\documentclass[10pt]{iopart}
\usepackage{epsfig}
\usepackage{rotating}
\usepackage{graphics}
\usepackage{graphicx}

\newcommand{\vsp}{\vspace*{3mm}}
\newcommand{\bra}{\langle}
\newcommand{\ket}{\rangle}
\newcommand{\kav}{{\langle k \rangle}}
\newcommand{\order}{{\cal O}}
\newcommand{\R}{{\rm I\!R}}
\newcommand{\bsigma}{{\mbox{\boldmath $\sigma$}}}
\newcommand{\btheta}{{\mbox{\boldmath $\theta$}}}
\newcommand{\bxi}{{\mbox{\boldmath $\xi$}}}
\newcommand{\bc}{\ensuremath{\mathbf{c}}}
\newcommand{\bk}{\ensuremath{\mathbf{k}}}
\newcommand{\by}{\ensuremath{\mathbf{y}}}
\newcommand{\atanh}{{\rm atanh}}
 \newcommand{\A}{{\mathcal A}}

\begin{document}

\title[Network resilience against attacks constrained by node removal cost]{Network resilience against intelligent attacks constrained by degree dependent node removal cost}

\author{A Annibale$^\dag$, ACC Coolen$^{\dag\diamond}$, and G Bianconi$^\ddag$}
\address{$\dag$ Department of Mathematics, King's College London,\\ The Strand,
London WC2R 2LS, United Kingdom}
\address{$\diamond$ Randall Division of Cell and Molecular Biophysics,
King's College London, New Hunt's House, London SE1 1UL, United Kingdom}
\address{$\ddag$ Northwestern University, Boston, USA
}

\begin{abstract}
We study the resilience of 
complex networks against attacks in which nodes 
are targeted intelligently, but where 
disabling a node has a cost to the attacker which depends on its degree. Attackers have  
to meet these costs with limited resources, which constrains their actions. 
A network's integrity is quantified in terms of the efficacy of the process that it supports.  
We calculate how the optimal attack strategy and the most attack-resistant network degree statistics
depend on the node removal cost function and the attack resources.
The resilience of networks against intelligent attacks is found to depend strongly on the 
node removal cost function faced  by the attacker. 
In particular, if node removal costs increase sufficiently fast
with the node degree, power law networks are found to be more resilient than 
Poissonian ones, even against optimised intelligent attacks. 
\end{abstract}

\ead{alessia.annibale@kcl.ac.uk,  ton.coolen@kcl.ac.uk, g.bianconi@neu.edu}

\section{Introduction}

In recent years there have been several studies into the resilience of complex networks against 
random failures and targeted attacks, in which a fraction of the nodes or of the bonds are removed.
It was found that scale-free networks   (with degree distributions that decay slowly via power laws, 
as in preferential attachment models) 
are more robust against random node removal
than Poissonnian (or Erd\"{os}-R\'{e}nyi) graphs, which may explain why many real-world
complex systems involve networks with power-law distributed
degrees. However, scale-free networks were found to be very vulnerable to
intelligent attackers  that target high-degree nodes \cite{AlbJeoBar00,CohEreBenHav01}. 
Against edge removal, 
Poissonian and power law networks turned out to produce similar responses 
\cite{MagLatGui09}. 
There are two reasons why we aim to study network resilience further. 
First, while the motivation behind such studies is that the networks provide the infrastructure
for some process (with interacting `agents' or processors occupying the nodes), 
and that process disruption is the true goal of an attacker,  
most authors measure the impact of attacks indirectly, via topological
properties that serve as proxies for the integrity of the process (e.g. the overall
connectivity and path-length statistics  \cite{AlbBar02}, or percolation characteristics \cite{AlbJeoBar00,CohEreBenHav01,NewStrWat01,CalNewStrWat00,CohEreBenHav00}).  
Here we seek to quantify the damage inflicted by attacks directly in terms of the process 
which the network is meant to support, similar to \cite{HasMen07}.
This requires solving stochastic processes on complex networks with arbitrary degree distributions, which is what statistical mechanics enable us to do. 
Our second and most important reason is that 
network resilience has so far been studied strictly in the context of random or intelligent removal of a fixed {\em fraction} 
of sites or bonds. This seems unrealistic. In most real-world scenarios (attacks on computer networks, viruses attacking cellular networks, etc) 
attacking a highly connected node demands more effort on behalf of the attacker than removing a weakly connected one. Similarly, any sensible defender of a network 
would devote more resources to the protection of `hubs' than to the protection of `outpost' nodes. 
The study of network resilience against attack or dilution calls for more appropriate and realistic definitions, that include 
the inevitable resource constraints faced by attackers and defenders alike. 

Turning to a formulation where attackers have {\it finite resources}, to be deployed intelligently 
when the cost of removing a network node depends on the degree of that node, changes 
the game drastically. It introduces a trade-off between the merit 
in terms of inflicted damage of targeting high-degree nodes versus
the disadvantage of associated cost (attacking many `hubs' may be inaffordable). 
One would like to know the maximum amount of damage that can be inflicted (by e.g. a virus to a biological network),  
given the limited resources available to the attacker (e.g. food, lifetime) 
and given the network's degree-dependent node removal costs. 
Similarly one would like to identify the most resilient network degree statistics to withstand an optimal attack.  
The answers to these questions  may aid our understanding of structural properties of biological (e.g. proteomic) signalling networks, 
where competition and natural selection act as driving forces towards attack resistance, but also to aid the design of attack-resistant 
synthetic real-world (e.g. communication) networks. 

Here we develop a framework for the study of network resilience that includes 
limited attack resources, degree-dependent node removal costs,  and resilience 
measures based on process integrity. We
consider two types of processes where structurally different interacting variables are placed on the nodes of
networks with arbitrary degree distributions: interacting Ising spins (where global order is ferromagnetic or of the spin-glass type), and  
coupled Kuramoto oscillators (where global order is measured by synchronization).
Both are solvable using finite connectivity replica theory, which enables us to quantify their integrity by the critical temperature of the ordered state. 
An attacker with finite resources seeks to destabilize these processes   by removing or disrupting selected network nodes 
using his knowledge of the network's degrees. The attacker is also allowed to disable nodes 
partially (with a proportional reduction in attack costs). 
We identify the most damaging attack strategy, given a network's degree distribution and given the 
degree dependence of the node removal costs and the attack resources available. 
We then determine the optimal network topology from the point of view of the defender, i.e. that degree distribution 
for which the integrity of the process is preserved best when attacked by a foe who employs the most damaging attack strategy. 
The optimal attack strategy and the optimally attack-resistant network topology are found to be universal across the types of microscopic
variables and types of global order considered.
As expected, the resilience of network processes against intelligent attacks depends strongly on the 
node removal cost function faced  by the attacker. 
Moreover, in sharp contrast to the traditional set-up where attackers are allowed to remove a fixed {\em fraction} of the nodes (and hence can simply target the `hubs'), 
we find that  if node removal costs increase sufficiently fast
with the node degree,  and if attackers have finite resources to meet these costs, power law networks are more resilient than 
Poissonian ones, even against optimised intelligent attacks. 

\section{Definitions}

\subsection{Processes, supporting networks, and constrained attack variables}

We study two systems in which interacting stochastic variables are placed on the $N$ nodes 
of a complex network. The network is defined via variables $c_{ij}\in\{0,1\}$, with $c_{ij}=1$ if and only if the nodes $i$ and $j$ are connected.
  We define $c_{ij}=c_{ji}$ and $c_{ii}=0$ for all $(i,j)$, and abbreviate $\bc=\{c_{ij}\}$. 
The first system (A) consist of $N$ Ising spins $\sigma_i\in\{-1,1\}$, in thermal
equilibrium, characterized by the following Hamiltonian
\begin{eqnarray}
{\rm A\!:}&&~~~~ H(\bsigma)=-\sum_{i<j}c_{ij}J_{ij}\xi_i\xi_j\sigma_i \sigma_j
\label{eq:Hamiltonian_A}
\end{eqnarray}
with $\bsigma=(\sigma_1,\ldots,\sigma_N) $.
The second system (B) consist of $N$ Kuramoto oscillators, with phases $\theta_i\in[-\pi,\pi]$, again in equilibrium but now with the Hamiltonian
\begin{eqnarray}
{\rm B\!:}&&~~~~ H(\btheta)=-\sum_{i<j}c_{ij}J_{ij}\xi_i\xi_j \cos(\theta_i-\theta_j)
\label{eq:Hamiltonian_B}
\end{eqnarray}
with $\btheta=(\theta_1,\ldots,\theta_N)$. 
The bonds $J_{ij}\in\R$ are drawn
randomly and independently from a distribution  $P(J)$. The variables $\xi_i\in\Xi\subseteq[0,1]$ 
in (\ref{eq:Hamiltonian_A},\ref{eq:Hamiltonian_B}) represent the impact of attacks, with $\xi_i=0$
 if node $i$ is removed completely and $\xi_i=1$ if it is left alone. We demand that $1\in\Xi$, so leaving a node intact is always an option, and 
 for simplicity we take $\Xi$ to be discrete and finite.
We define the node degrees
$k_i(\bc)=\sum_{j}c_{ij}$, so the degree
distribution and the average connectivity of $\bc$ are $p(k|\bc)=N^{-1}\sum_i \delta_{k,k_i(\bc)}$ and 
$\bra k\ket=\sum_{k\geq 0} kp(k|\bc)$, respectively.

We assume that the cost to the attacker of setting $\xi_i=\xi$ at a node of degree $k_i=k$ is $\psi(\xi,k)\geq 0$,
where $\psi(1,k)=0$ and $\partial \psi(\xi,k)/\partial\xi\leq 0$ for all $\xi\in\Xi$. 
If attackers have limited resources they can only disrupt a subset of the nodes, since 
 the $\{\xi_i\}$ will now be subject to a constraint of the form $\sum_i \psi(\xi_i,k_i)\leq C$. A natural choice for $\psi$ is
 \begin{eqnarray}
 \psi(\xi,k)=\kappa(1\!-\!\xi)\phi(k)
 \label{eq:attack_cost}
 \end{eqnarray}
where $\phi(k)$ is a non-decreasing function,
with  $\phi(0)=0$ and $\phi(k\!>\!0)>0$. 
The attack cost for a node increases with the number of
links to/from it;  
disconnected nodes can be attacked for free. The normalization factor $\kappa$ is chosen such that the resource constraint takes the simple form 
$N^{-1}\sum_i   \psi(\xi_i,k_i)\leq 1$.
The attacker is assumed to act intelligently, using knowledge of the network's degrees, so the degrees $\{k_i\}$ and the attack variables $\{\xi_i\}$ will generally be correlated.
Finally we draw the network $\bc$ randomly from a maximum-entropy ensemble defined by a probability distribution in which
 the degrees are constrained to take prescribed values $\bk=(k_1,\ldots,k_N)$:
\begin{eqnarray}
{\rm Prob}(\bc)&=& Z^{-1}[\bk] \prod_i \delta_{k_i,k_i(\bc)},~~~~~~Z[\bk]=\sum_{\bc}\prod_i \delta_{k_i,k_i(\bc)}
\label{eq:connectivity}
 \end{eqnarray}
 We  abbreviate $p(\xi,k)=N^{-1}\sum_i\delta_{\xi_i,\xi}\delta_{k_i,k}$, and define $q(\xi|k)$ via $p(\xi,k)=q(\xi|k)p(k)$. 
 The resource constraint on the
 attack variables then translates into $\sum_{\xi k}\psi(\xi,k)q(\xi|k)p(k)\leq 1$.
 The attacker is assumed to know the degree sequence $\bk$ of the network to be attacked,
 and can adapt accordingly the conditional likelihood $q(\xi|k)$ to maximize the impact of his actions;
 $q(\xi|k)$ constitutes his attack strategy. 
 The realistic regime is that where $\psi(\xi,k)$ obeys $\sum_{k}\psi(0,k)p(k)>1$, so that the trivial `destroy-all'
 attack strategy $q(\xi|k)=\delta_{\xi,0}~ \forall k$  is not feasible (i.e. too costly).
 
 \subsection{Quantifying process integrity and optimal attack and defence strategies}

With each process (\ref{eq:Hamiltonian_A},\ref{eq:Hamiltonian_B}) running on the network and each associated ordered phase 
(ferromagnetic, spin-glass, or synchronized) corresponds a critical temperature $T_c$,
which will for large $N$ depend on the network and attack characteristics $\bk$ and $\bxi$ only via 
$q(\xi|k)$ and $p(k)$. The larger $T_c$,  the more robust is the ordered phase against 
local noise, so we can quantify the integrity of the process by the value of $T_c[p,q]$.
The attacker wants to destroy the ordered phase of the process, whereas the defender seeks to protect it.
This allows us to give precise definitions for the optimal attack strategy
 and the optimally resistant degree distribution in terms of process integrity. The optimal attack strategy $q^\star[p]$ is
 the conditional distribution $q(\xi|k)$  for which $T_c[p,q]$ is minimal, given the degree distribution $p$ and given the 
 resource constraint:
 \begin{eqnarray}
 q^\star[p]&=&{\rm argmin}_{\{q,\sum_{\xi k}\psi(\xi,k)q(\xi|k)p(k)\leq 1\}} T_c[p,q]
 \label{eq:optimal_attack}
 \end{eqnarray}
 The optimal (most resistant) degree distribution $p^\star$
 to be chosen by the defender, given the average connectivity $c$ (finite network
 resources) and attack cost function $\psi$ is then that $p(k)$ which subsequently maximizes
 this $q$-minimized critical temperature:
 \begin{eqnarray}
 p^\star&=&{\rm argmax}_{\{p,\sum_k kp(k)=c\}}T_c[p,q^\star[p]]
 \nonumber
 \\
 &
= &
 {\rm argmax}_{\{p,\sum_k kp(k)=c\}}{\rm argmin}_{\{q,\sum_{\xi k}\psi(\xi,k)q(\xi|k)p(k)\leq 1\}} T_c[p,q]
 \nonumber
 \\&&
  \label{eq:optimal_defense}
  \end{eqnarray}
  The end result is a situation where the defender, by choosing an appropriate degree distribution, 
  maintains the highest achievable critical temperature $T_c[p^\star,q^\star[p^\star]]$, given he
   is subjected to the most damaging attack. However,
  within this scenario one could in fact  ask many more interesting questions, such as what would be the effect of  misinformation, a situation where a defender
  optimizes the network on the basis of an anticipated attack $q^\star[p]$ (so he chooses degree distribution $p^\star$)
  but is then faced with an attack with strategy  $q^\prime\neq q^\star[p]$, so that the actual critical temperature is
  $T_c[p^\star,q^\prime]$.

  We see that the problem of identifying the optimal attack and defense strategies (\ref{eq:optimal_attack},\ref{eq:optimal_defense})
  splits automatically into two distinct parts. The first part is 
  calculating the critical temperature(s) $T_c[p,q]$ of the relevant phases. This is done 
  by evaluating for the systems (\ref{eq:Hamiltonian_A},\ref{eq:Hamiltonian_B}) the asymptotic disorder-averaged
 free energy per spin $\overline{f}$, from which one extracts the phase diagrams  for systems on typical graphs from  (\ref{eq:connectivity}):
  \begin{eqnarray}
 \overline{f}_{\!A}&=&-\lim_{N\to\infty}\frac{1}{\beta N}\overline{\log \sum_{\bsigma}\rme^{-\beta H(\bsigma)}}
 \label{eq:fA}
 \\
 \overline{f}_{\!B}&=&-\lim_{N\to\infty}\frac{1}{\beta N}\overline{\log \int_{-\pi}^\pi\!\rmd\btheta~\rme^{-\beta H(\btheta)}}
 \label{eq:fB}
 \end{eqnarray}
 in which $\beta=T^{-1}$ (where $T$ denotes the temperature), and where $\overline{\cdots}$ denotes averaging over the disorder in the problem, viz. the randomly drawn graphs with statistics
 (\ref{eq:connectivity}) and  the random bonds $\{J_{ij}\}$. The calculation of (\ref{eq:fA}) and (\ref{eq:fB})
 is done with the finite connectivity
 replica method, based on the identity $\overline{\log Z}=\lim_{n\to 0}n^{-1}\log \overline{Z^n}$, and details are relegated to \ref{app:spins} and \ref{app:oscillators} in order not to disrupt the flow of the paper.
 The second part of the problem, to be tackled once the formulae for $T_c[p,q]$ have been derived (which, expectedly and fortunately, 
 turn out to be simple and very similar across models and ordered phases), is to carry out the constrained optimizations in 
 (\ref{eq:optimal_attack},\ref{eq:optimal_defense}), by a combination of analytical and numerical techniques.

\section{The process integrity measure}
\label{sec:attack_defense}

We show in the appendices of this paper that the critical temperatures $T_c[p,q]$ for the emergence of global (F or SG) order, given we choose the bond distribution
$P(J)=\frac{1}{2}(1\!+\!\eta)\delta(J\!-\!J_0)+\frac{1}{2}(1\!-\!\eta)\delta(J\!+\!J_0)$ (with $J_0\geq 0$), 
follow for both Ising spins and coupled oscillators 
from formulae of the following form:
\begin{eqnarray}
\hspace*{-20mm}
{\rm F}:&~~~& \lambda^{(1)}_{\rm max}(\beta)=1,~~~~~\lambda^{(1)}\!:~~{\rm eigenvalues~of~} M^{(1)}_{\xi\xi^\prime}(\beta)=\eta K(\beta J_0\xi\xi^\prime)\gamma(\xi^\prime)
\label{eq:PtoFgeneral}
\\
\hspace*{-20mm}
{\rm SG}:&~~~& \lambda^{(2)}_{\rm max}(\beta)=1,~~~~~\lambda^{(2)}\!:~~{\rm eigenvalues~of~} M^{(2)}_{\xi\xi^\prime}(\beta)=K^2(\beta J_0\xi\xi^\prime)\gamma(\xi^\prime)
\label{eq:PtoSGgeneral}
\end{eqnarray}
in which $\beta=1/T$ and 
\begin{eqnarray}
\gamma(\xi)&=& \bra k\ket^{-1}\sum_k q(\xi|k)p(k)k(k-1),
\label{eq:gamma_xi}
\end{eqnarray}
 Here $K(z)=\tanh(z)$ for interacting Ising spins and
$K(z)=I_1(z)/I_0(z)$ for coupled oscillators. In both cases $K(-z)=-K(z)$, $\frac{\rmd}{\rmd z} K(z)\geq 0$, and $\lim_{z\to\infty}K(z)=1$.
There is no F phase if $\eta\leq 0$, so we take $\eta>0$ from now on.
The structure of the above formulae is in agreement with results from percolation theory and spreading phenomena, which show that the threshold characterizing the percolation 
transition or an epidemic outbreak in a network depends on the ratio $\bra k^2\ket/\bra k\ket$ of the first two moments of its degree distribution 
\cite{AlbJeoBar00,NewStrWat01,CalNewStrWat00,CohEreBenHav00,CohEreBenHav01,PasVes01a,PasVes01b,LloMay01}. 
The approach followed here is closer to the envisaged picture of interacting agents or processors on network nodes, and  has the benefit 
of applying to the whole interval $\Xi=[0,1]$, as opposed to $\Xi=\{0,1\}$ which can be accessed by percolation theory.

\subsection{Tests and bounds for critical temperatures}

Before any attack one  has $\xi\in\{1\}$, so $\gamma(\xi)=\gamma(1)=\bra k^2\ket/\bra k\ket-1$ and the above formulae would have reproduced the known results for the unperturbed system, viz.
\begin{eqnarray}
{\rm F}:&~~~& \eta K(\beta J_0)[\bra k^2\ket/\bra k\ket-1]=1
\label{eq:PtoFstandard}
\\
{\rm SG}:&~~~& K^2(\beta J_0)[\bra k^2\ket/\bra k\ket-1]=1
\label{eq:PtoSGstandard}
\end{eqnarray}
Another simple test is to consider $\Xi=\{0,1\}$.
Here each node is either unaffected or removed completely, leaving a new network identical to an unperturbed network
 as described by (\ref{eq:PtoFstandard},\ref{eq:PtoSGstandard}), but with reduced size $N^\prime=\sum_i \xi_i$, and with degrees $k_i^\prime=\sum_j c_{ij}\xi_j$. 
 We would find, in the case of random attacks $q(\xi|k)=\zeta\delta_{\xi,0}+(1-\zeta)\delta_{\xi,1}$:
 \begin{eqnarray}
 \hspace*{-10mm}
 \bra k\ket^\prime &=&\lim_{N\to\infty}\frac{1}{(1\!-\!\zeta)N}\sum_{ij}\xi_i c_{ij}\xi_j=(1\!-\!\zeta)\bra k\ket
 \\[-1mm]
 \hspace*{-10mm}
 \bra k^2\ket^\prime &=&\lim_{N\to\infty}\frac{1}{(1\!-\!\zeta)N}\sum_{ij\ell}\xi_i c_{ij}c_{i\ell}\xi_j\xi_\ell=(1\!-\!\zeta)^2\bra k^2\ket+\zeta(1\!-\!\zeta)\bra k\ket
\end{eqnarray}
giving the following transparent  formulae for the post-attack transition points:
\begin{eqnarray}
{\rm F}:&~~~& \eta K(\beta J_0)(1\!-\!\zeta)[\bra k^2\ket/\bra k\ket-1]=1
\label{eq:PtoFtrivial}
\\
{\rm SG}:&~~~& K^2(\beta J_0)(1\!-\!\zeta)[\bra k^2\ket/\bra k\ket-1]=1
\label{eq:PtoSGtrivial}
\end{eqnarray}
If, alternatively, we apply to this scenario the result (\ref{eq:PtoFgeneral},\ref{eq:PtoSGgeneral}), we find 
$\gamma(\xi)=q(\xi)[\bra k^2\ket/\bra k\ket-\!1]$ and $K(\beta J_0\xi\xi^\prime)=K(\beta J_0)\delta_{\xi,1}\delta_{\xi^\prime,1}$,
and the relevant matrices reduce to $M^{(1)}_{\xi\xi^\prime}(\beta)=\eta K(\beta J_0)(1\!-\!\zeta)[\bra k^2\ket/\bra k\ket\!-\!1]\delta_{\xi,1}\delta_{\xi^\prime,1}$ and $M^{(2)}_{\xi\xi^\prime}(\beta)=K^2(\beta J_0)(1\!-\!\zeta)[\bra k^2\ket/\bra k\ket\!-\!1]\delta_{\xi,1}\delta_{\xi^\prime,1}$. One solves the eigenvalue problems trivially, and indeed recovers
(\ref{eq:PtoFtrivial}, \ref{eq:PtoSGtrivial}). 
A final trivial test is to consider $q(\xi|k)=\delta_{\xi,\xi_0}$, where $\xi_0\in(0,1)$, an attack 
equivalent to replacing $J_0\to \xi_0^2J_0$. Upon substituting this choice 
into (\ref{eq:PtoFgeneral},\ref{eq:PtoSGgeneral}) one confirms, via $\gamma(\xi)=\delta_{\xi,\xi_0}[\bra k^2\ket/\bra k\ket\!-\!1]$, that our general theory indeed reduces to (\ref{eq:PtoFstandard},\ref{eq:PtoSGstandard}) with the correctly reduced coupling strength.

Solving the eigenvalue problems (\ref{eq:PtoFgeneral},\ref{eq:PtoSGgeneral}) analytically is not always possible, but eigenvalue bounds
are obtained easily. Our matrices are of the form
$M_{\xi\xi^\prime}=L(\xi\xi^\prime)\gamma(\xi^\prime)$, where $L(u)=\eta K(\beta J_0 u)$ for the F transition (so $L(u)$ is anti-symmetric)
and  $L(u)=K^2(\beta J_0 u)$ for the SG transition (so $L(u)$ is symmetric), and where $\gamma(\xi)\geq 0$ for all $\xi$.
We symmetrize the eigenvalue problem $\lambda x(\xi)=\sum_{\xi^\prime}M_{\xi\xi^\prime}x(\xi^\prime)$ by defining
$y(\xi)=x(\xi)\sqrt{\gamma(\xi)}$, giving   $\lambda y(\xi)=\sum_{\xi^\prime}[\sqrt{\gamma(\xi)}L(\xi\xi^\prime)\sqrt{\gamma(\xi^\prime)}]y(\xi^\prime)$.
This implies that
\begin{eqnarray}
\lambda_{\rm max}&=& \max_{\by}\frac{\sum_{\xi\xi^\prime} y(\xi)\sqrt{\gamma(\xi)}L(\xi\xi^\prime)\sqrt{\gamma(\xi^\prime)}y(\xi^\prime)}{
\sum_\xi y^2(\xi)}
\end{eqnarray}
 which can be simplified  to
\begin{eqnarray}
\lambda_{\rm max}&=& \max_{\by}\frac{\sum_{\xi\xi^\prime} y(\xi)\sqrt{\gamma(\xi)}L(|\xi\xi^\prime|)\sqrt{\gamma(\xi^\prime)}y(\xi^\prime)}{
\sum_\xi y^2(\xi)}
\label{eq:maxy}
\end{eqnarray}
Variational arguments can now be applied in order to get lower bounds.
In particular, upon substituting $y(\xi)=\delta_{\xi,\hat{\xi}}$ and varying $\hat{\xi}$ one derives the statement
\begin{eqnarray}
\lambda_{\rm max}&\geq & \max_{\xi}\Big\{ \gamma(\xi)L(\xi^2)\Big\}
\end{eqnarray}
To find upper bounds, we use the fact that the maximum in (\ref{eq:maxy}) will have $y(\xi)\geq 0$ for all $\xi$.
We then use the inequalities $L(|u|)\leq \alpha\eta\beta J_0|u|$ (for F) and $L(|u|)\leq (\alpha\beta J_0)^2|u|^2$ (for SG),
where $\alpha=1$ for Ising spins and $\alpha=\frac{1}{2}$ for coupled oscillators,  to get
\begin{eqnarray}
\lambda_{\rm max}^{(1)}&\leq & \alpha\eta\beta J_0
\max_{\by}\Big\{
\frac{\sum_{\xi\xi^\prime} y(\xi)\gamma^{\frac{1}{2}}(\xi)|\xi||\xi^\prime|\gamma^{\frac{1}{2}}(\xi^\prime)y(\xi^\prime)}{
\sum_\xi y^2(\xi)}\Big\}
\\
\lambda_{\rm max}^{(2)}&\leq & (\alpha\beta J_0)^2
\max_{\by}\Big\{
\frac{\sum_{\xi\xi^\prime} y(\xi)\gamma^{\frac{1}{2}}(\xi)|\xi|^2|\xi^\prime|^2\gamma^{\frac{1}{2}}(\xi^\prime)y(\xi^\prime)}{
\sum_\xi y^2(\xi)}\Big\}
\end{eqnarray}
The last two maxima are calculated easily, leading us to
\begin{eqnarray}
{\rm F:}&~~~& T_c[p,q]~\leq~ \eta \alpha J_0 \sum_\xi \xi^2\gamma(\xi)
\label{eq:TcF}
\\[-1mm]
{\rm SG:}&~~~& T_c[p,q]~\leq~ \alpha J_0 \Big(\sum_\xi \xi^4\gamma(\xi)\Big)^{\!\frac{1}{2}}
\label{eq:TcSG}
\end{eqnarray}

\subsection{Explicit simple form for a process integrity measure $\Gamma[p,q]$}

The inequalities (\ref{eq:TcF},\ref{eq:TcSG})  become equalities for large $c$, 
where the critical temperatures diverge and hence $\beta\to 0$ in (\ref{eq:PtoFstandard},
\ref{eq:PtoSGstandard}); the right-hand sides of (\ref{eq:TcF},\ref{eq:TcSG}) then become the true integrity measures of the process. 
Moreover, for certain natural choices of the set $\Xi$ the latter statement  is in fact true for {\em any} connectivity $c$. For instance,  
 if $\Xi\subseteq \{0,1\}$ (all nodes are either fully disabled or left alone) one may use $K(\beta J_0\xi\xi^\prime)=\xi\xi^\prime K(\beta J_0)$ to diagonalize the matrices in  (\ref{eq:PtoFgeneral},\ref{eq:PtoSGgeneral}) and find
\begin{eqnarray}
{\rm F:}&~~~& \frac{1}{K(J_0/T_c[p,q])}=\eta\sum_\xi \xi^2\gamma(\xi)
\\
{\rm SG:}&~~~& \frac{1}{K(J_0/T_c[p,q])}= \Big(\sum_\xi \xi^4\gamma(\xi)\Big)^{\frac{1}{2}}
\end{eqnarray}
which reveals that the critical temperatures are monotonically increasing functions of the sums
$\sum_\xi \xi^2\gamma(\xi)$ for F-type order and $\sum_\xi \xi^4\gamma(\xi)$ for SG-type order (for $\Xi=\{0,1\}$ the two sums are in fact identical). 
In view of these properties, and in view of the minor differences between  the F and SG cases,  
 in the remainder of this study we adopt the quantity $\sum_\xi \xi^2\gamma(\xi)$ as our integrity measure, giving 
\begin{eqnarray}
\Gamma[p,q]&=& \frac{1}{\bra k\ket}\sum_{\xi k}\xi^{2} q(\xi|k)p(k)k(k-1)
\label{eq:Gamma_bound}
\end{eqnarray}
We define the set of relevant degrees $k$ as $S=\{k>1|~p(k)>0\}$. 
The optimal attack strategy is then the choice $q^\star[p]$ which solves the following optimization problem:
\begin{eqnarray}
{\rm minimize:}&~~& \Gamma[p,q]
\label{eq:optimize}
\\[1mm]
{\rm subject~to:}&~~&
q(\xi|k)\geq 0~\forall (\xi,k),~~~\sum_{\xi\in\Xi}q(\xi|k)=1~\forall k\in S
\\[-1mm]
&&\sum_{\xi \in\Xi}\sum_{k\in S}(1\!-\!\xi)\phi(k) q(\xi|k)p(k)\leq \kappa^{-1}
\label{eq:constraints}
\end{eqnarray}
To avoid trivial pathologies we  assume that $\exists k\ge 2$ with $p(k)>0$ (if untrue we would not have an ordered state in the first place, 
as it would have given $T_c[p,q]=0$), and that $\sum_k p(k)k(k-1)<\infty$ (if untrue there would not be a finite critical temperature before the attack).
Clearly $q^\star(\xi|k)=\delta_{\xi 1}$ for $k\notin S$; any other choice would sacrifice attack resources without benefit. 
The best defense against optimal attacks is the choice for the degree distribution $p(k)$ such that the above
minimum over $q$ is maximized.

\subsection{Bounds on the process integrity measure}

To judge the quality of attack strategies it will prove useful to have bounds on the value $\Gamma[p,q^\star[p]]$ corresponding to the optimal attack $q^\star[p]$.
An upper bound is easily obtained by inspecting the result of non-intelligent random attacks of the type $q(\xi|k)=(1-Q)\delta_{\xi,0}+Q\delta_{\xi,1}$, with $0\leq Q\leq 1$:
\begin{eqnarray}
&&\Gamma[p,q]=[\bra k^2\ket/\bra k\ket-1] Q
\label{Gamma}
\\
&& Q\geq 1-1/\kappa\bra \phi(k) \ket
\end{eqnarray}
The sharpest bound of this form follows when seeking equality in the last line, giving\footnote{Note that $\kappa\bra \phi(k) \ket> 1$ due to our earlier ruling out of the trivial attack strategy $q(\xi|k)=\delta_{\xi,0}$.}:
\begin{eqnarray}
\Gamma[p,q^\star[p]]&~\leq~ &\Gamma[p,q^\star_{\rm random~I}]= \Big(1-
\frac{1}{\kappa \bra \phi(k) \ket}\Big)
\bra k(k\!-\!1)\ket/\bra k\ket
\label{eq:upperboundGamma_I}
\end{eqnarray}
If $\Xi=\{0,1\}$ then (\ref{eq:upperboundGamma_I}) is the best possible upper bound based on random attacks.
If $\Xi=[0,1]$ we can improve upon (\ref{eq:upperboundGamma_I}) by investigating random attacks of the form $q(\xi|k)=\delta[\xi-\hat{\xi}]$. The optimal choice turns out to be $\hat{\xi}=1\!-\!1/\kappa\bra \phi\ket$, giving
\begin{eqnarray}
\Gamma[p,q^\star[p]]&~\leq~ &\Gamma[p,q^\star_{\rm random~II}]=\Big(1-
\frac{1}{\kappa \bra \phi(k) \ket}\Big)^{\!2}
\bra k(k\!-\!1)\ket/\bra k\ket
\label{eq:upperboundGamma_II}
\end{eqnarray}
To find lower bounds for $\Gamma[p,q^\star[p]]$ we first define modified probabilities $\pi(k)\in[0,1]$:
\begin{eqnarray}
\pi(k) &=& \frac{\sum_{\xi\in\Xi}(1\!-\!\xi)q(\xi|k)\phi(k) p(k)}{\big\bra\sum_{\xi\in\Xi}(1\!-\!\xi)q(\xi|k)\phi(k)\big\ket}
\label{eq:define_pi}
\end{eqnarray}
with associated averages written as $\bra \ldots\ket_\pi$.
Note that the denominator of (\ref{eq:define_pi}) is bounded from above by $\kappa^{-1}$, via the resource constraint.   We can now write
\begin{eqnarray}
\hspace*{-15mm}
\Gamma[p,q]&=& \frac{\bra k(k\!-\!1)\ket}{\bra k\ket} -\frac{1}{\bra k\ket}\sum_k p(k)k(k\!-\!1)\sum_{\xi\in\Xi}(1\!-\!\xi^2)q(\xi|k)
\nonumber
\\
\hspace*{-15mm}
&=& \frac{\bra k(k\!-\!1)\ket}{\bra k\ket} -\frac{1}{\bra k\ket}
\Big\bra\sum_{\xi\in\Xi}(1\!-\!\xi)q(\xi|k)\phi(k)\Big\ket\Big\bra \frac{k(k\!-\!1)}{\phi(k)}\frac{\sum_{\xi\in\Xi}(1\!-\!\xi^2)q(\xi|k)}{
\sum_{\xi\in\Xi}(1\!-\!\xi)q(\xi|k)}\Big\ket_{\!\pi}
\!\!\!\!
\nonumber
\\
\hspace*{-15mm}
&\geq & \frac{\bra k(k\!-\!1)\ket}{\bra k\ket} -\frac{1}{\kappa \bra k\ket}
\Big\bra \frac{k(k-1)}{\phi(k)}\frac{\sum_{\xi\in\Xi}(1\!-\!\xi^2)q(\xi|k)}{
\sum_{\xi\in\Xi}(1\!-\!\xi)q(\xi|k)}\Big\ket_\pi
\nonumber
\\
\hspace*{-15mm}
&\geq & \frac{\bra k(k\!-\!1)\ket}{\bra k\ket} -\frac{C_\Xi}{\kappa\bra k\ket}~\Big\bra
\frac{k(k-1)}{\phi(k)}\Big\ket_\pi
\end{eqnarray}
in which the factor $C_\Xi\geq 0$ depends only on the choice made for the value set $\Xi$:
\begin{eqnarray}
\hspace*{-10mm}
C_\Xi&=& \max_{w}\Big\{\frac{1\!-\!\bra \xi^2\ket_w}{1\!-\!\bra \xi\ket_w}\Big\},~~{\rm with}~~
\bra f(\xi) \ket_w=\sum_{\xi\in\Xi}w(\xi)f(\xi)~~{\rm and}~~\sum_{\xi\in \Xi}w(\xi)=1
\nonumber
\\[-2mm]
\hspace*{-10mm}
\end{eqnarray}
One easily proves using $\Xi\subseteq [0,1]$ that $C_\Xi\in[1,2]$, that $C_{\{0,1\}}=1$, and that $C_{[0,1]}=2$.
We conclude, in combination with (\ref{eq:upperboundGamma_I},\ref{eq:upperboundGamma_II}), that
\begin{eqnarray}
\hspace*{-20mm}
\Xi=\{0,1\}:&~~~&
1 \!-\!\frac{1}{\kappa\bra k(k\!-\!1)\ket}\Big\bra
\frac{k(k\!-\!1)}{\phi(k)}\Big\ket_{\!\!\pi} \leq~\frac{\bra k\ket \Gamma[p,q^\star[p]]}{\bra k(k\!-\!1)\ket} ~\leq~
1\!-\!\frac{1}{\kappa\bra \phi(k) \ket}
\label{eq:GammaboundsI}
\\
\hspace*{-20mm}
\Xi=[0,1]:&~~~&
1\! -\!\frac{2}{\kappa \bra k(k\!-\!1)\ket} \Big\bra
\frac{k(k\!-\!1)}{\phi(k)}\Big\ket_{\!\!\pi} \leq~ \frac{\bra k\ket \Gamma[p,q^\star[p]]}{\bra k(k\!-\!1)\ket} ~\leq~
\Big(1\!-\!
\frac{1}{\kappa \bra \phi(k) \ket}\Big)^{\!2}
\label{eq:GammaboundsII}
\end{eqnarray}
The lower bounds are satisfied with equality if the attack resources are exhausted and if $\bra (1-\xi^2)\ket_q=\bra (1-\xi)\ket_q$ for each
$q(\xi|k)$ with $k\in S$; the last condition is always met if $\Xi=\{0,1\}$. However, the lower bounds 
still depend on the attack strategy via the measure $\pi$.
From (\ref{eq:GammaboundsI},\ref{eq:GammaboundsII}) and the general property $\Gamma[p,q]\geq 0$, which follows from the definition
of $\Gamma[p,q]$, we finally obtain the strategy-independent bounds
\begin{eqnarray}
\hspace*{-20mm}
\Xi=\{0,1\}:&~~~
{\rm max}\Big\{0,~
1 \!-\!\frac{R/\kappa}{\bra k(k\!-\!1)\ket}\Big\} &~\leq~\frac{\bra k\ket \Gamma[p,q^\star[p]]}{\bra k(k\!-\!1)\ket} ~\leq~
1
\!-\!\frac{1}{\kappa\bra \phi(k) \ket}
\label{eq:GammaBoundsI}
\\
\hspace*{-20mm}
\Xi=[0,1]:&~~~
{\rm max}\Big\{0,~1 \!-\!\frac{2R/\kappa}{\bra k(k\!-\!1)\ket}\Big\} &~\leq~ \frac{\bra k\ket \Gamma[p,q^\star[p]]}{\bra k(k\!-\!1)\ket} ~\leq~
\Big(1\!-\!
\frac{1}{\kappa \bra \phi(k) \ket}\Big)^{\!2}
\label{eq:GammaBoundsII}
\end{eqnarray}
with
\begin{eqnarray}
R&=&{\rm max}_{k\in S}\big\{k(k\!-\!1)/\phi(k)\big\}
\label{eq:R}
\end{eqnarray}
The latter bounds reveal immediately two distinct
situations where it is not possible for {\em any} intelligent attack to improve on the
damage done by random attacks: the case $\phi(k)=k(k-1)$ for all $k\in S$ (here the benefit of degree knowledge exactly balances the cost to the attacker of using it), 
and the case of regular random graphs, viz.
$p(k)=\delta_{k,\bra k\ket}$, where there is no degree knowledge to be exploited in the first place.

\section{Optimal attack and optimal defense for $\Xi=\{0,1\}$}

The attacker's objective is to minimize $\Gamma[p,q]$.  We have seen that for $\Xi=\{0,1\}$, where nodes are either fully disabled or left alone and 
$C_{\{0,1\}}=1$, the 
 lower bound in  (\ref{eq:GammaBoundsI}) could in principle be realized. This will serve as an efficient guide in finding $q^\star[p]$. 
 Attack strategies for $\Xi=\{0,1\}$ are of the form $q(\xi|k)=q(0|k) \delta_{\xi,0}+[1-q(0|k)]\delta_{\xi,1}$, 
 so we need to determine $q(0|k)$ for all $k\in S$.

\subsection{Construction of the optimal attack strategy}

We first define the attacker's `target' degree set $\A\subseteq S$, with $R$ as defined in (\ref{eq:R}):
\begin{eqnarray}
\A&=&\{k\in S|~k(k-1)/\phi(k)=R\}
\label{eq:targetset}
\end{eqnarray}
The inequality $\bra k(k\!-\!1)/\phi(k)\ket_\pi\leq R$ used in the final step of our derivation of (\ref{eq:GammaBoundsI}) 
is satisfied with {\em equality} only if $\pi(k)=0$ for all $k\notin\A$. 
According to (\ref{eq:define_pi}) this requires $q(\xi|k)=\delta_{\xi,1}$ for all $k\notin\A$.
The only remaining requirement for satisfying the lower bound in (\ref{eq:GammaBoundsI}) is that we satisfy
the resource constraint with equality. Hence the set of optimal attack strategies is defined strictly by the following
demands:
\begin{eqnarray}
\forall k\notin\A:&~~&q(0|k)=0
\label{eq:optimal_condition1}
\\
\forall k\in\A:&~~&q(0|k)\in[0,1],~~~
\sum_{k\in\A}q(0|k)\phi(k)p(k)=1/\kappa
\label{eq:optimal_condition2}
\end{eqnarray}
It is straightforward to verify directly, using $\phi(k)=k(k\!-\!1)/R$ for all $k\in\A$, that strategies satisfying 
these conditions indeed give the lowest possible value for $\Gamma[p,q]$ according to our bounds, and satisfy the resource constraint with equality.
By construction, the set $\A$ cannot be empty. 

At this stage in our argument we must distinguish between two distinct cases. 
In the first case the attacker need not look beyond nodes in the target set $\A$ (\ref{eq:targetset}), since removing those will already exhaust or exceed his resources;
he will simply remove as many of those as can be afforded. In the second case the removal of all nodes in $\A$ does not exhaust the attack resources, 
and new target sets need to be identified: 
\begin{itemize}
\item
The target set $\A$ is exhausting, $\sum_{k\in\A}\phi(k)p(k)\geq 1/\kappa$:
\\[2mm]
Here it is immediately clear that optimal attacks will indeed exist, i.e. the conditions (\ref{eq:optimal_condition1},\ref{eq:optimal_condition2}) 
can be met. Only nodes from $\A$ will be removed. 
If
there is at least one $k^\star\in\A$ with $\phi(k^\star)p(k^\star)\geq 1/\kappa$, the attacker can simply execute
\begin{eqnarray}
k^\star={\rm argmax}_{k\in\A}\{\phi(k)p(k)\}\\
q(0|k^\star)=1/\kappa p(k^\star)\phi(k^\star),~~~~~~
\forall k\neq k^\star:~q(0|k)=0
\end{eqnarray}
If instead $\phi(k)p(k)< 1/\kappa$ for all $k\in\A$ 
there is no target degree in $\A$ which would on its own exhaust the attacker's resources. The attacker will first remove all  
nodes with degree $k_1^\star={\rm argmax}_{k\in\A}\{\phi(k)p(k)\}$ by setting 
$q(0|k_1^\star)=1$. He will next direct attention to the reduced set $\A/\{k^\star_1\}$ and 
remove nodes with degree $k_2^\star={\rm argmax}_{k\in\A/\{k_1^\star\}}\{\phi(k)p(k)\}$, etc, 
until the resources are exhausted. At the end of this iterative process  the attacker will have removed a sequence of degrees $\{k_1^\star,\cdots,k_L^\star\}\subseteq\A$
(where nodes with degree $k_{L}^\star$ will generally be only partially removed, as allowed by remaining resources).  
In words: 
the attacker first determines the target set $\A$ of those degrees with $p(k)>0$ for which the ratio $k(k-1)/\phi(k)$ is maximal.
He then ranks the degrees in $\A$ according to the value of $\phi(k)p(k)$ and proceeds to remove degrees iteratively 
according to this ranking until his resources are exhausted. 
This strategy will always lead to 
$\bra k(k-1)/\phi(k)\ket_\pi =R$, and satisfy the lower bound in (\ref{eq:GammaBoundsI}) with equality.
\vsp

\item
The target set $\A$ is non-exhausting, $\sum_{k\in\A}\phi(k)p(k)< 1/\kappa$:
\\[2mm]
Here the attacker can afford to remove completely all degrees in the set $\A$, 
but setting $q(0|k)=1$ for all $k\in\A$ does not exhaust his resources. 
He should subsequently direct
attention to those nodes in the reduced set $S/\A$ for which the ratio $k(k-1)/\phi(k)$ is maximal, and so on.
The result is again an iteration, at the end of which the attacker will have removed a set of degrees $\{k_1^\star,\cdots,k_L^\star\}\supset\A$
(where nodes with degree $k_{L}^\star$ will generally be only partially removed).  
In this case $\bra k(k-1)/\phi(k)\ket<R$ and the lower bound in (\ref{eq:GammaBoundsI}) is no longer satisfied with inequality;
however, this does not imply that the strategy is non-optimal, since it might be that the bound is no longer tight.
Here it is therefore difficult to prove rigorously that the identified strategy always constitutes the optimal attack, but it is the logical continuation of 
the optimal attack identified earlier and its optimality is consistent with numerical experiments (to be shown later).
\end{itemize}
\vsp

\noindent
We can combine both cases above in a transparent iterative attack protocol.  
We define at each step $\ell$: the target set $\A_\ell$, the set $S_\ell$ of nodes that have not yet been targeted, 
and the resource remainder $\Delta_\ell=\kappa^{-1}-\sum_k q(0|k)\phi(k)p(k)$.
The process is initialized according to $S_0=S$ and $\Delta_0=\kappa^{-1}$, and starts with $q(0|k)=0$ for all $k$. It is iterated until $\Delta_\ell=0$, according to

\begin{description}
\item[~~~~~{\em step 1:}] calculate new ratio $R_\ell=\max_{k\in S_{\ell-1}}\{k(k\!-\!1)/\phi(k)\}$
\item[~~~~~{\em step 2:}] identify target set $\A_\ell=\{k\in S_{\ell-1}|~k(k\!-\!1)/\phi(k)=R_\ell\}$
\item[~~~~~{\em step 3:}] choose (any) $k_\ell^\star\in\A_\ell$ for which $\phi(k_\ell^\star)p(k_\ell^\star)={\rm max}_{k\in \A_\ell}\{\phi(k)p(k)\}$ 
\item[~~~~~{\em step 4:}] check whether attack resources can be exhausted:
\begin{eqnarray*}
\hspace*{-10mm}
\phi(k_\ell^\star)p(k_\ell^\star)\geq \Delta_{\ell-1}:&~~~& {\rm yes,}\\[-1mm]
\hspace*{-10mm}&& {\rm remove~as~many~degree~}k_\ell^\star~{\rm nodes~as~possible}
\\
\hspace*{-10mm}
&& {\rm set}~q(0|k_\ell^\star)=\Delta_{\ell-1}/ \phi(k_\ell^\star)p(k_\ell^\star)
\\
\hspace*{-10mm}
&& \Delta_\ell=0,~{\rm attack~terminates}
\\[2mm]
\hspace*{-10mm}
\phi(k_\ell^\star)p(k_\ell^\star)<\Delta_{\ell-1}:&~~~& {\rm no,}
\\[-1mm]
\hspace*{-10mm} && {\rm remove~all~degree}~k_\ell^\star~{\rm nodes}
\\
\hspace*{-10mm}&&
{\rm set}~q(0|k_\ell^\star)=1
\end{eqnarray*}

\item[~~~~~{\em step 5:}] 
define $S_\ell=S_{\ell-1}/k_\ell^\star~$ and $~\Delta_\ell=\Delta_{\ell-1}-
\phi(k_\ell^\star)p(k_\ell^\star)q(0|k_\ell^\star)$
\end{description}
\vsp

\noindent
Always the end result is a sequence $\{k_1^\star,\ldots,k_{L-1}^\star\}$ of target degrees that are fully 
removed, possibly supplemented by a further degree $k_L^\star$ of which a fraction will be removed (to exhaust fully the 
 attack resources).

\subsection{Properties of the optimal attack strategy}

We next evaluate the impact of the above attack strategy $q^\star[p]$ on our process integrity measure. 
 We define the set $\A^\star=\{k_1^\star,\ldots,k_{L-1}^\star\}$  of fully removed degrees, and write 
$k_L^\star$ simply as $k^\star$.
The post-attack value of the process integrity measure will be
 \begin{eqnarray}
 \hspace*{-10mm}
\Gamma[p,q^\star[p]]&=&\frac{\bra k(k\!-\!1)\ket}{\bra k\ket}- \frac{1}{\bra k\ket}\!\sum_{k\in\A^\star}p(k)k(k\!-\!1)
-\frac{1}{\bra k\ket}q(0|k^\star)k^\star(k^\star\!\!-\!1)p(k^\star)
\nonumber
\\
 \hspace*{-10mm}
&=& \frac{\bra k(k\!-\!1)\ket}{\bra k\ket}- \frac{1}{\bra k\ket}\sum_{k\in\A^\star}p(k)k(k\!-\!1)
-\frac{\Delta_{L-1}}{\bra k\ket}\frac{k^\star(k^\star\!\!-\!1)}{ \phi(k^\star)}
\end{eqnarray}
The attack $q^\star[p]$ exhausts all resources, so $\Delta_{L-1}=\kappa^{-1}-\sum_{k\in\A^\star}\phi(k)p(k)$. Hence
 \begin{eqnarray}
\Gamma[p,q^\star[p]]
&=&\frac{\bra k(k\!-\!1)\ket}{\bra k\ket}- \frac{1}{\bra k\ket}\!\sum_{k\in\A^\star}p(k)\phi(k)\Big[\frac{k(k\!-\!1)}{\phi(k)}
-\frac{k^\star(k^\star\!-\!1)}{ \phi(k^\star)}\Big]
\nonumber
\\&&\hspace*{30mm}
-\frac{k^\star(k^\star\!\!-\!1)}{\kappa\bra k\ket \phi(k^\star)}
\label{eq:finalGammaXi01}
\end{eqnarray}
Since by definition $k(k\!-\!1)/\phi(k)>k^\star(k^\star\!-\!1)/\phi(k^\star)$ for all $k\in\A^\star$, both the second 
and the third term of (\ref{eq:finalGammaXi01}) are strictly non-positive.

The result (\ref{eq:finalGammaXi01}) can be compared to that of the most damaging random attack (where no degree information is used), 
viz. to (\ref{eq:upperboundGamma_I}).
The benefit $\Delta\Gamma=\Gamma[p,q^\star[p]]-\Gamma[p,q^\star_{\rm random}]$ to the attacker of using optimal intelligent attacks as opposed to optimal 
random attacks then takes the form
\begin{eqnarray}
\hspace*{-17mm}
\Delta\Gamma &=&
- \!\frac{1}{\bra k\ket}\!\sum_{k\in\A^\star}p(k)\phi(k)
\Big[\frac{k(k\!-\!1)}{\phi(k)}\!-\!\frac{k^\star(k^\star\!\!-\!1)}{ \phi(k^\star)}\Big]
-\frac{1}{\kappa\bra k\ket}\Big[
\frac{k^\star(k^\star\!\!-\!1)}{\phi(k^\star)}\!-\!\frac{\bra k(k\!-\!1)\ket}{ \bra \phi(k) \ket}
\Big]
\nonumber
\\
\hspace*{-17mm}
&=& 
- \frac{1}{\bra k\ket}\!\sum_{k\in\A^\star}p(k)\phi(k)\Big[\frac{k(k\!-\!1)}{\phi(k)}-\frac{k^\star(k^\star\!\!-\!1)}{ \phi(k^\star)}\Big]
\nonumber
\\
\hspace*{-17mm}
&&\hspace*{30mm}
- \frac{1}{\kappa\bra k\ket\,\bra\phi(k)\ket}\sum_k p(k)\phi(k)\Big[\frac{k^\star(k^\star\!\!-\!1)}{ \phi(k^\star)}-\frac{k(k\!-\!1)}{\phi(k)}\Big]
\nonumber
\\
\hspace*{-17mm}
&=& 
- \frac{1}{\bra k\ket}\!\sum_{k\in\A^\star}p(k)\phi(k)\Big[\frac{k(k\!-\!1)}{\phi(k)}-\frac{k^\star(k^\star\!\!-\!1)}{ \phi(k^\star)}\Big]
\left(1-\frac{1}{\kappa\bra\phi(k)\ket}\right)
\nonumber
\\
\hspace*{-17mm}
&& \hspace*{20mm}
+ \frac{1}{\kappa\bra k\ket\,\bra\phi(k)\ket}\sum_{k\notin\A^\star} p(k)\phi(k)\Big[\frac{k(k\!-\!1)}{\phi(k)}-\frac{k^\star(k^\star\!-\!1)}{\phi(k^\star)}\Big]
\end{eqnarray}
Since the set $\A^\star\subseteq S$
is constructed specifically from those degrees for which $k(k-1)/\phi(k)$ is maximal,
and since $\bra\phi(k)\ket>\kappa^{-1}$, both terms of $\Delta\Gamma$ are strictly non-positive. One will thus generally have $\Delta\Gamma<0$.
Again we also recognize the two special cases where there will be no gain in intelligent attacks, namely  $\phi(k)=k(k-1)$ (with any degree distribution),  
and $p(k)=\delta_{k,\bra k\ket}$ (with any cost function $\phi(k)$).

In terms of the dependence of our results on the cost function $\phi(k)$ it is clear that everything evolves around 
the dependence on $k$ of the ratio $\phi(k)/k(k-1)$. This ratio represents for each $k$ the balance between the cost of removing 
degree-$k$ nodes versus the benefits in terms of damage achieved.   
If for simplicity we choose $\phi(k)=k^\zeta(k-1)$, then for $\zeta<1$ the intelligent attack will be to take out first the nodes with the
{\em largest} degrees $k\in S$ that can be removed without violating the resource constraint (i.e. the `greedy' attack strategy is optimal), 
whereas for $\zeta>1$ the intelligent attack will target first the nodes with the {\em smallest} degrees $k\in S$ that can be removed without violating the constraint
(here attacking hubs is too expensive to be efficient). Furthermore, it is not at all a priori clear what would be the most resistant degree distribution against such attacks in the presence of resource constraints. Naively one could perhaps have expected that for small $\zeta$ (where the attacker will target hubs) the best strategy for the defender could be to 
choose a narrowly distributed degree distribution, so there are no hubs to be exploited. Interestingly, we will see below that that is not the case, and the optimal degree distribution can be more subtle.

\section{Numerical results}

\subsection{General methods}

\begin{table}[t]
\centering\vsp
\begin{tabular}{lrrrrrrl}
Species & $N~~$ & $\kav~$ & $k_{\mbox{\rm max}}$ & Method & Reference
\\
\hline
C. Elegans & 3512 & 3.72 & 524 & Y2H & \cite{PPINs1}~~~~\\
C. Jejuni & 1324 & 17.52 & 207 & Y2H & \cite{PPINs2}~~~~\\
E. Coli & 2457 & 7.05 & 641 & PMS & \cite{PPINs5}~~~~\\
H. Sapiens & 9306 & 7.53 & 247 & HPRD & \cite{PPINs4}~~~~\\
S. Cerevisiae & 3241 & 2.69 & 279 & Y2H & \cite{PPINs3}~~~~\\
\hline\\[-3mm]
\end{tabular}
\caption{Size $N$, average connectivity $\kav$, maximum degree $k_{\mbox{max}}$, experimental detection method or source, and reference for the biological (protei interaction) network data sets used in our numerical experiments. 
The detection methods or sources are abbreviated as follows:
Y2H, Yeast two-hybrid; PMS, Purification-Mass spectrometry; HPRD, the human protein reference database.
}
\label{tab:properties}
\end{table}

In this section we illustrate, apply and extend  via numerical experimentation the results derived above.
We determine by numerical maximization the most resistant degree distribution against optimal intelligent attacks, and we
compare for typical biological networks 
the effects of optimal intelligent attacks in terms of process integrity against optimized random attacks and against the bounds established earlier.
The biological networks used are experimentally determined protein interaction networks (PINs) 
of different species, namely C. Elegans, C. Jejuni, E. Coli, H. Sapiens and S. Cerevisiae; 
see Table~\ref{tab:properties} for characteristics  and references.
For each biological network we also generate several synthetic alternatives with
the same size $N$ and average connectivity $\kav$ as the biological one, but with different degree distributions:
Poissonian, the optimally resistant degree distribution, or a
distribution generated via preferential attachment with a fat 
tail similar to the biological network.
In all cases we choose node attack cost 
functions of the form $\phi(k)=k^{\zeta}(k-1)$, with $\zeta=0,1,2$, and we set  the attack resource limit 
 to $\kappa^{-1}=\bra\phi(k)\ket_{\mathcal P}/q$, where $q>1$ is a control parameter and 
the average $\bra\dots\ket_{\mathcal P}$
is calculated over the Poissonian distribution ${\mathcal P}(k)=\kav^k e^{\kav}/k!$
For each PIN and each synthetically generated counterpart  
the average cost function $\bra \phi(k)\ket$  is found to be always larger than or equal to 
the one calculated over the equivalent Poissonian distribution 
(if $\zeta=0$, where $\phi(k)=k-1$, equality of course holds trivially for {\em all} distributions since 
they share by construction the value of $\bra k\ket$). 
This ensures that for  $q>1$ the attacker's resources will in all our experiments 
be in the relevant regime $\bra\phi(k)\ket > \kappa^{-1}$.
The optimally attack resistant networks are found via a stochastic graph dynamics,
starting from a biological protein interaction network,  
in which at each step a bond is selected at random and is moved to 
another location if this move increases the post-attack integrity measure $\Gamma[p,q^\star[p]]$. 
Bond relocations are the minimal moves that preserve the average degree of the network.
After each move, the optimal attack strategy $q^\star[p]$ defined in the previous section 
is applied to the new network. 
In order to prevent the graph dynamics from getting stuck in suboptimal configurations, 
we allow initially for groups of bonds to be moved, and as the algorithm proceeds the  
size of these groups  is reduced, in the spirit of  \cite{KueLinPoe98}.

\begin{figure}[t]
\hspace*{-13mm}
\begin{picture}(200,190)
\put(150,170){\includegraphics[width=23.7\unitlength,angle=-90]{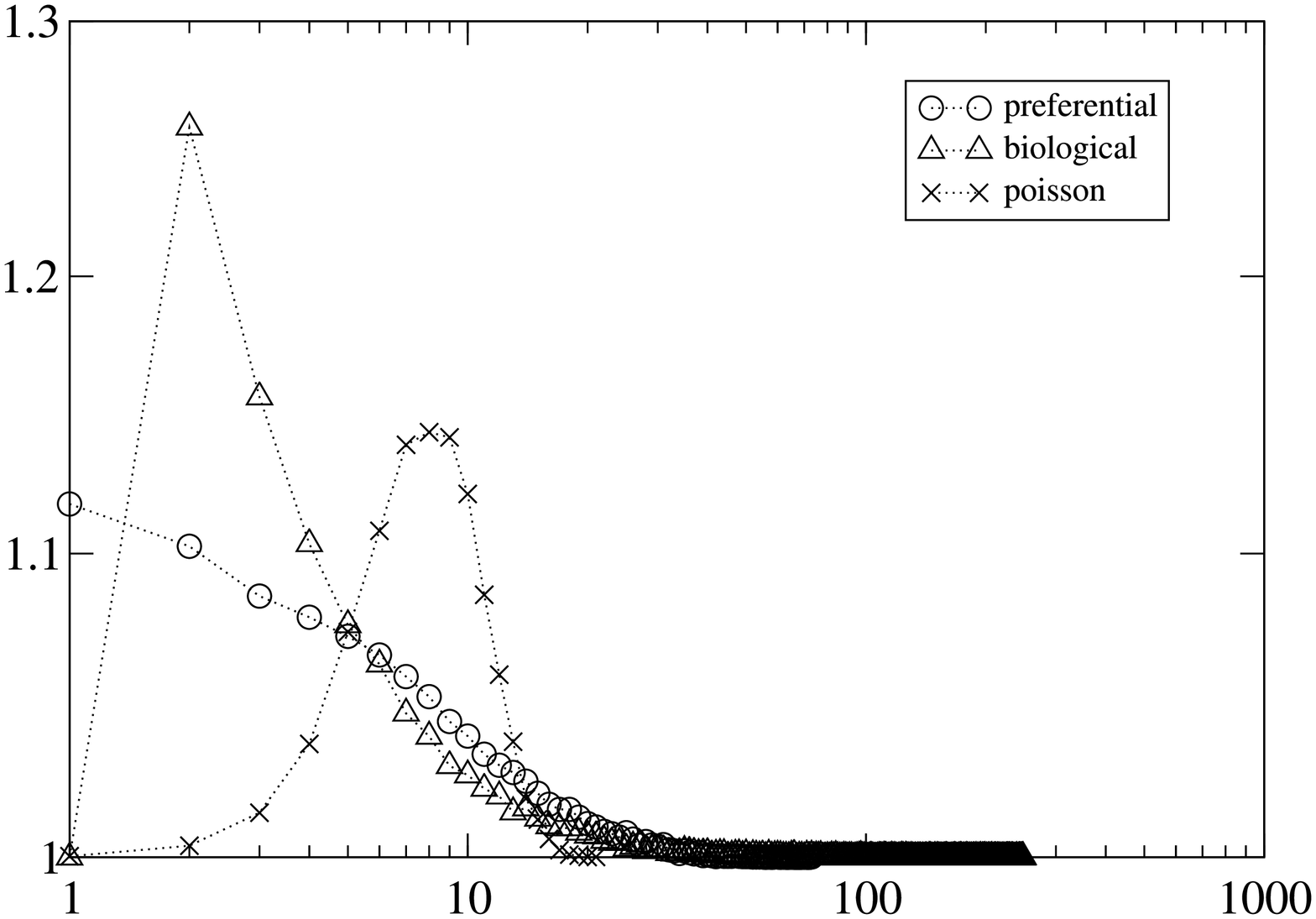}}
\put(-17,90){\rotatebox{90}{$p(k)+1$}}
\put(100,17){$k+1$}
\put(365,170){\includegraphics[width=25\unitlength,angle=-90]{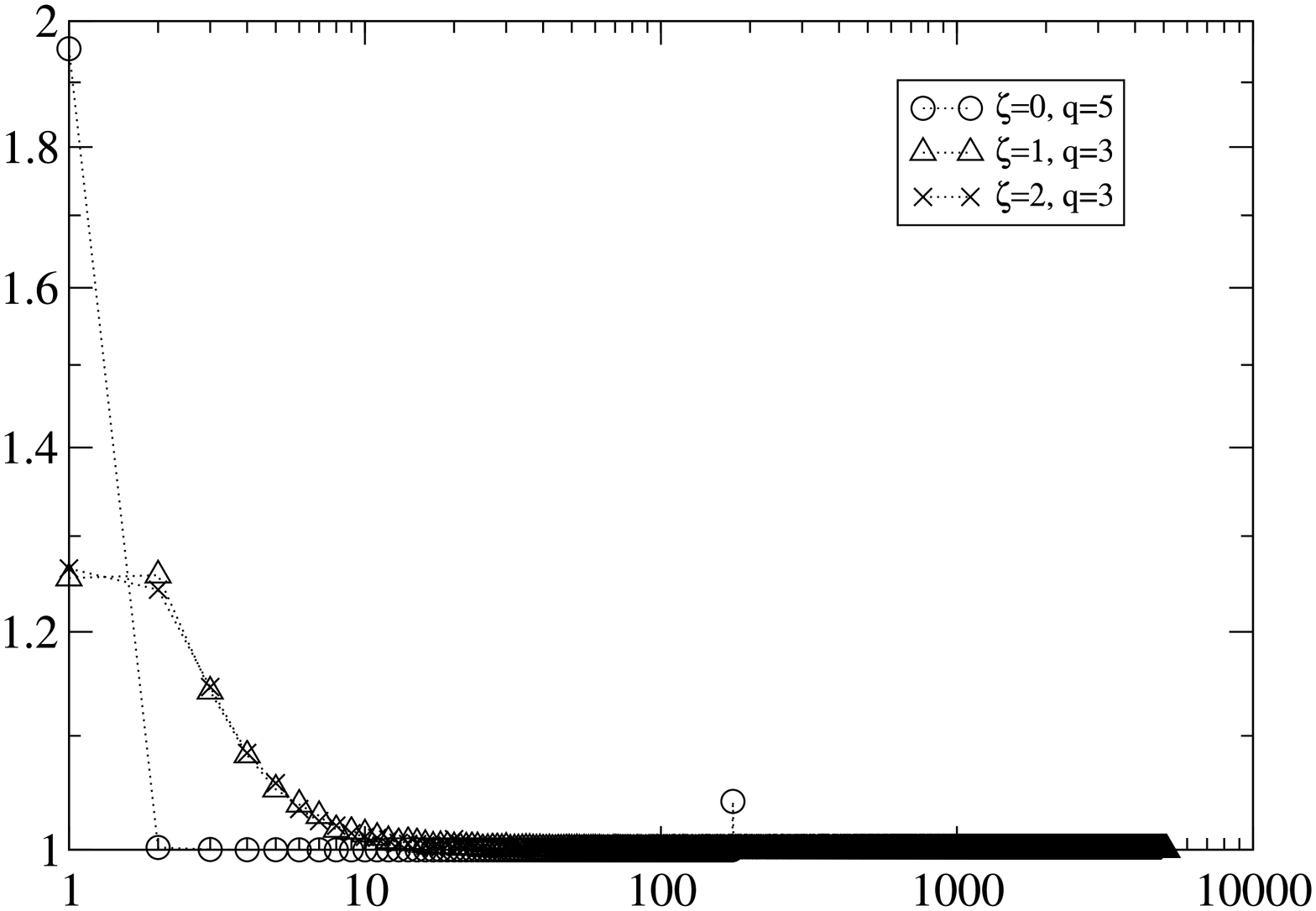}}
\put(315,17){$k+1$}
\end{picture}
\vspace*{-5mm}
\caption{Left: log-log plot of the degree distribution of the H. Sapiens PIN, 
 synthetically generated networks with the 
same size $N$ and average connectivity $\kav$ as H. Sapiens PIN but with Poissonian and preferential attachment 
degree distributions.
Right: log-log plot of the  degree distribution of the network that has the same size and average connectivity as the H. Sapiens PIN, 
but that has been constructed to be optimally resistant 
against optimal intelligent attacks, given the node removal cost function $\phi(k)=k^\zeta (k-1)$, with $\zeta=0,1,2$, 
and given available attack resources, characterized by $\kappa^{-1}=\frac{1}{q}\bra\phi(k)\ket_{\mathcal P}$, with $q=3,5$ (see legend).
We observe that upon decreasing $\zeta$, where the optimal attack strategy starts targeting the high-degree nodes, 
 the optimally resistant degree distribution changes takes a binary form, describing a module of high-degree nodes in a sea of unconnected nodes.  
}
\label{fig:degree_distribution}
\end{figure}

\subsection{Degree statistics before and after network optimization}

Figure \ref{fig:degree_distribution} shows the results of applying the above procedures  to the  
H. Sapiens PIN, for $\zeta=0,1,2$ and $q=3,5$. 
For $\zeta=1,2$ (where it is not advantageous to the attacker to target high-degree nodes) the optimally resistant 
degree distribution $p^\star[q]$ is seen to exhibit a smooth dependence on the degree $k$. 
For $\zeta=0$, where the degree dependence of node removal costs is modest and
 the optimal attack strategy is to target high degree nodes, one could expect the optimal network to 
 become regular, in order to disallow attackers to benefit from degree information. Instead, we observe an entirely different solution. 
Here, the optimal defender produces as many hubs as possible, so that the attacker is unable to remove all of these. The result is a distribution of the form 
\begin{eqnarray}
p^\star(k)&=& (1-\frac{\kav}{K})\delta_{k,0}+\frac{\kav}{K}\delta_{k,K},~~~~~~K\geq \bra k\ket
\label{eq:core}
\end{eqnarray}
with $K=192$ for a network with size and average connectivity identical to  
the H. Sapiens PIN. 
In a situation where attackers can and will target nodes with maximal degree first, 
it appears that the optimal defender chooses a network with a ``modular'' configuration, with a core of nodes 
highly connected to each other, in a sea of 
disconnected nodes. 
The attacker is prevented by resource limitations from 
removing more than a (tiny) fraction of the core.
The strategy of the optimal defender is to sacrifice a few highly connected nodes to save many. 
\begin{figure}[t]
\hspace*{-25mm}
\begin{picture}(200,430)
\put(200,400){\includegraphics[width=37\unitlength,angle=-90]{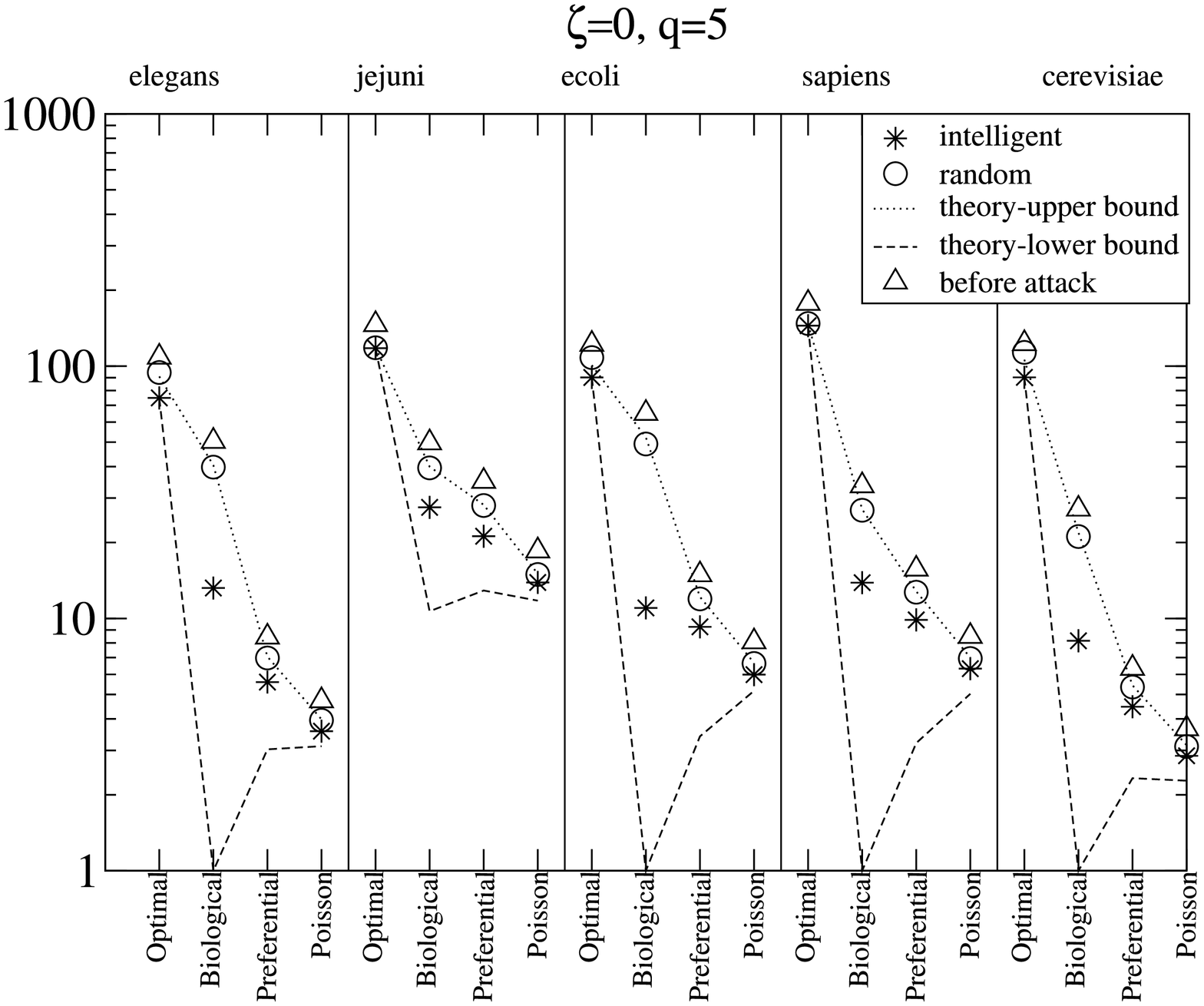}}
\put(30,310){\rotatebox{90}{$\Gamma[p,q]+1$}}
\put(430,400){\includegraphics[width=37\unitlength,angle=-90]{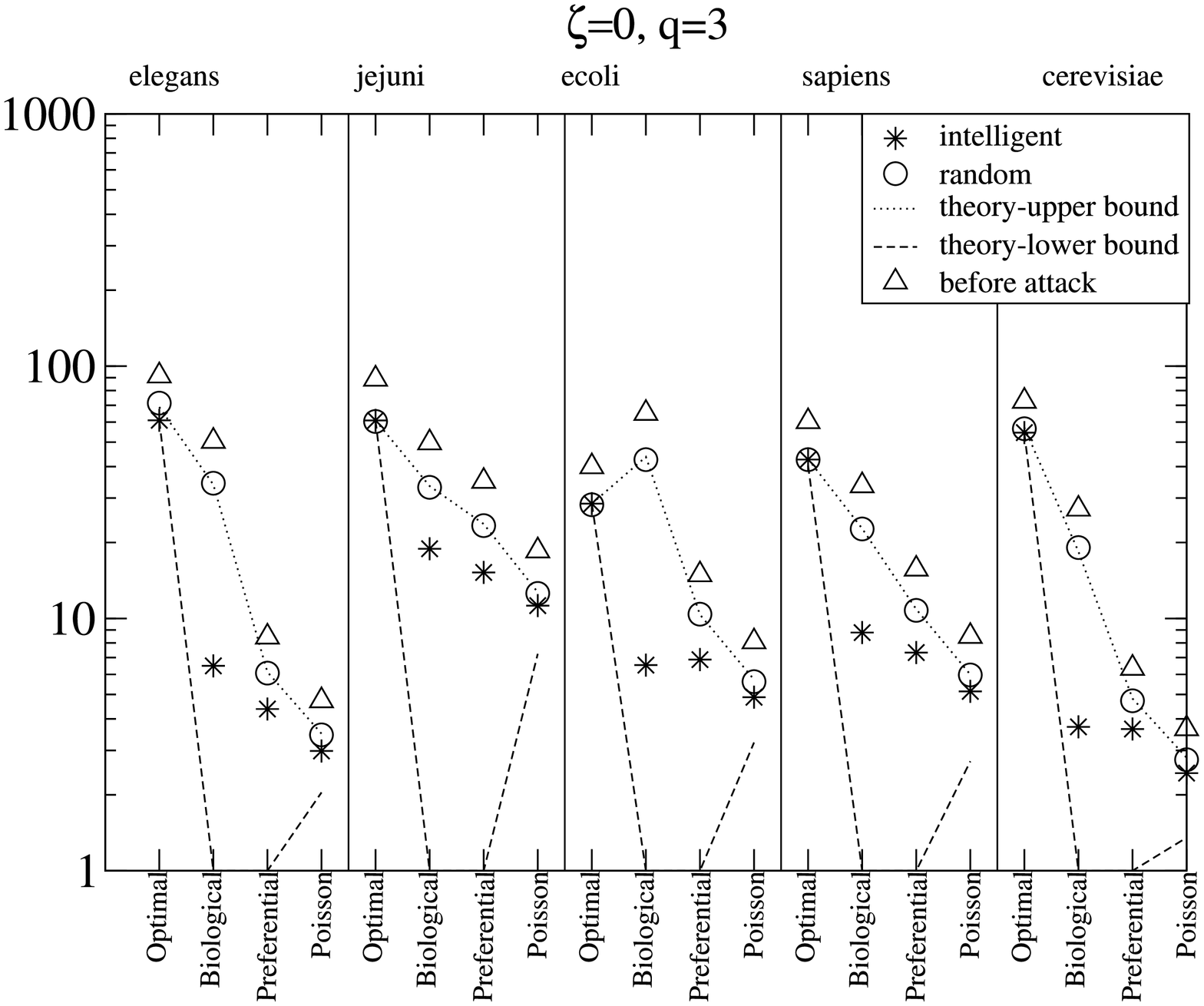}}
\put(200,190){\includegraphics[width=37\unitlength,angle=-90]{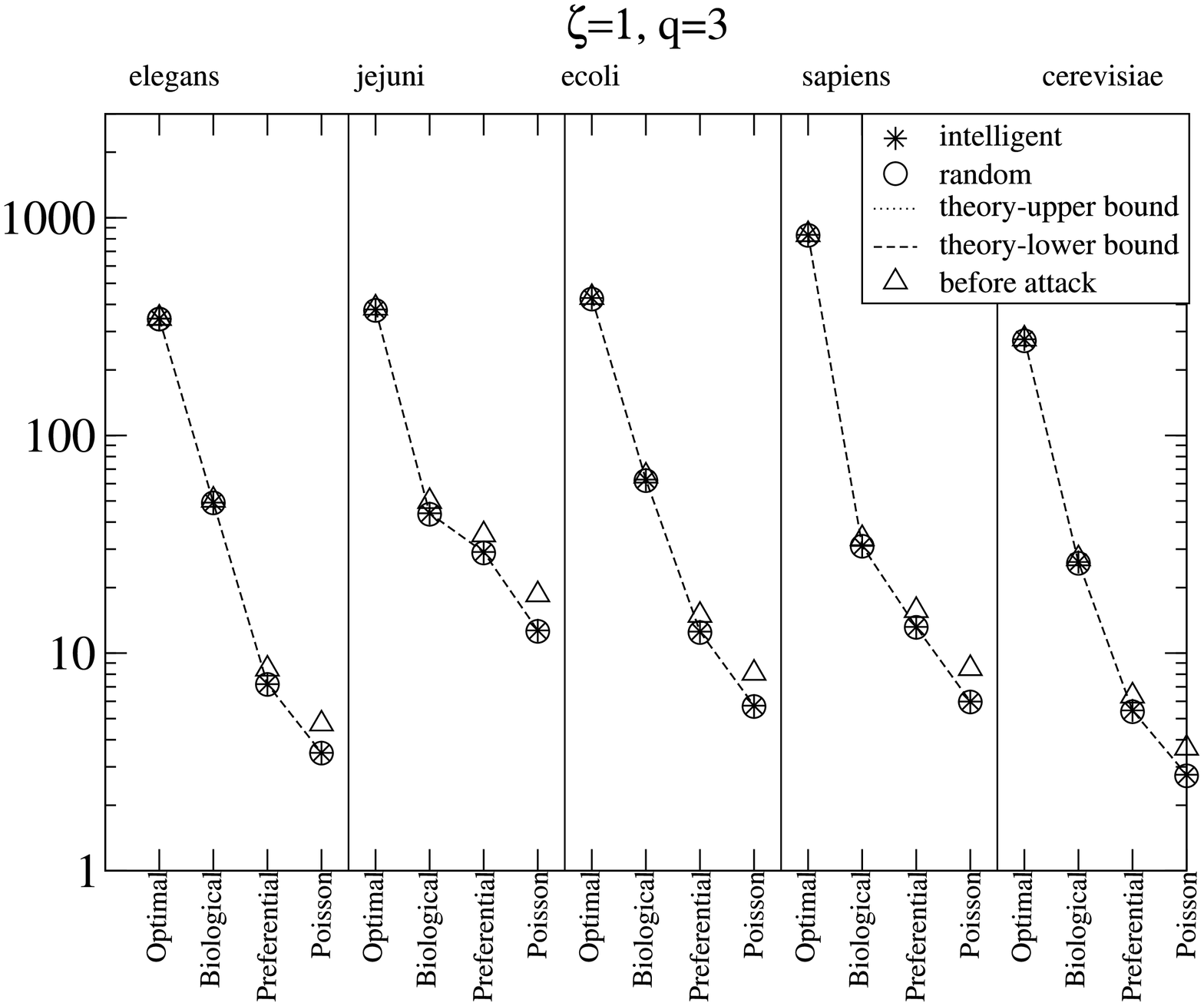}}
\put(30,90){\rotatebox{90}{$\Gamma[p,q]+1$}}
\put(430,190){\includegraphics[width=37\unitlength,angle=-90]{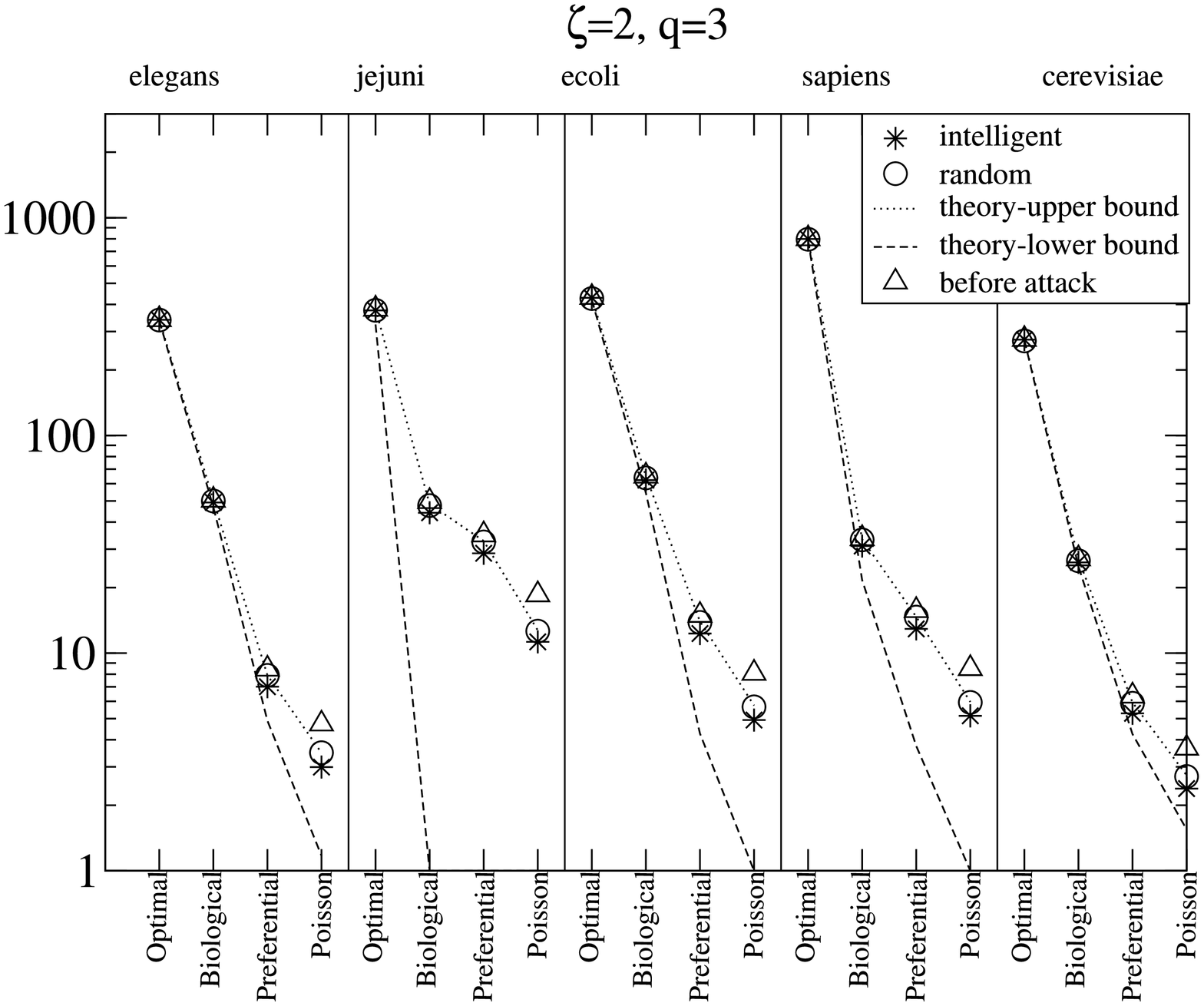}}
\end{picture}
\vspace*{-4mm}
\caption{Values of the process integrity measure $\Gamma[p,q]$ before attack ({\footnotesize $\triangle$}), after optimal intelligent attacks ({\large $*$}), and after optimal 
random attacks ({\large $\circ$}), 
for different networks. The specific networks
considered are experimentally determined PINs of different species 
(C. Elegans, C. Jejuni, E. Coli, H. Sapiens and S. Cerevisiae) and their synthetically generated 
counterparts with the same size and average connectivity, but different degree distributions
(Poissoninan, preferential attachment and optimally attack resistant degree distribution following the attack $q^\star[p]$).
The node attack cost function is $\phi(k)=k^{\zeta}(k-1)$ and the 
available attack resources are characterized by $\kappa^{-1}=\bra\phi(k)\ket_{\mathcal P}/q$,
with $\zeta=0,1,2$ and $q=3,5$ (see legends).
The theoretical upper and lower bounds (\ref{eq:GammaBoundsI}) are shown  as dotted and dashed lines, respectively.
All results consistently reproduce the
built-in order $\Gamma_{\rm before}\geq \Gamma_{\rm random}\geq \Gamma_{\rm intelligent}$ (viz. {\footnotesize $\triangle$}$\geq${\large $\circ$}$\geq${\large $*$}).
Furthermore, the network realizations are consistently ranked, with the optimally resistant network (as expected) always outperforming the others, but with 
also the biological and preferential attachment network outperforming their Poissonnian counterparts.
In fact,  the degree of resistance of the optimally resistant network is quite remarkable.
}
\label{fig:Gamma_species}
\end{figure}

For distributions of the form  (\ref{eq:core})  the attack resources can be exhausted, and the optimal attack is $q^\star(0|k)=\delta_{k,K}~K/ \kappa\phi(K)\bra k\ket$. 
This  results in
\begin{eqnarray}
\Gamma[p,q^\star[p]]&=& K-1
-K^{1-\zeta}/\kappa \bra k\ket
\end{eqnarray}
Insertion into (\ref{eq:GammaBoundsI})  shows that both bounds are now satisfied with equality. 
For $\zeta>0$ the defender would wish to choose $K$ as large as possible, but for a finite network there is a limit.
There are just $N\bra k\ket/K$ connected nodes; if each of these is to have $K$ neighbours we must demand 
$N\bra k\ket/K-1\geq K$, i.e. $K\leq K_c=\sqrt{N\bra k\ket}+\order(N^0)$. The same result follows from general entropic arguments 
\cite{AnnCooFerKleFra09,BiaCooVic08}. For large $N$ the number of graphs 
with degree distribution $p(k)$ equals $\exp[NS]$ where $S=\frac{1}{2}\bra k\ket[\log(N/\kav)+1]-\sum_k p(k)\log[p(k)/\pi(k)]$, with $\pi(k)=e^{-\bra k\ket}\bra k\ket^k/k!$.
 For (\ref{eq:core}) one obtains
\begin{eqnarray}
S
&=&\frac{1}{2}\bra k\ket\Big\{
1-\log\Big(\frac{K^2}{N\bra k\ket}\Big)\Big\}+\order(K^{-1}\log K)
\end{eqnarray}
Again we obtain the cut-off point $K\leq K_c\approx \sqrt{N\bra k\ket}$ for graphs with (\ref{eq:core}) to exist.
The value $K=192$ found numerically, see Fig. \ref{fig:degree_distribution},  is consistent with this bound (for H. Sapiens one has $K_c=264$), but not identical to it. 
This is expected to reflect finite size corrections to our theory, and the fact that the theory requires all relevant $k$ to be finite relative to $N$, whereas close to $K_c$ 
one has $k=\order(\sqrt{N})$.  

Note, however, that the distribution chosen by the optimal defender is not always exactly of the form (\ref{eq:core}). 
In some situations (depending on the amount of resources available to the attacker) the peak at $k=0$ is not strictly $\delta$-shaped, 
so that $K$ is no longer subjected to the previously identified cut-off $K_c$, and one indeed observes the second
peak to move to higher values of $K$ (albeit with a reduced height).
This results in a bimodal distribution with a $\delta$-peak at some $K>K_c$ and a broader peak at $k=0$,
corresponding to a strongly disassortative network configuration (for the notion of assortativity see e.g. \cite{Newman02}) 
where a small number of hubs are connected with an extremely large number of low degree nodes 
(reminescent of results derived in \cite{PauTanHavSta04}).
For this later distribution, as was the case with the bimodal distribution (\ref{eq:core}) with two strictly $\delta$-shaped peaks, 
the attacker will again exhaust his resources upon removal of just a tiny fraction, $q(0|K)=1/\kappa\phi(K)p(K)$, of hubs.

The actual distribution $p^{\star}(k)$ selected by the optimal defender 
when the attacker is bound, by resource limitations, 
to play the strategy $q^{\star}(0|k)=\delta_{k,K}\,1/\kappa\phi(k)p(k)$, is the one which maximizes the minimal integrity
measure 
\begin{eqnarray}
\Gamma(p,q^\star[p])&=&\frac{1}{\kav}\left(\bra k(k-1)\ket-\sum_k p(k)k(k-1)q(0|k)\right) 
\\&=&\frac{1}{\kav}\left(\bra k(k-1)\ket-\kappa^{-1} K^{1-\zeta}\right) 
\end{eqnarray}
achieved by the attacker.
It is clear that the shape of the optimally resistant distribution will depend on the interplay between $\bra k^2\ket$ 
and $K$, which is controlled by the resource limit $\kappa^{-1}$. For $\zeta=0$ numerical studies show that for 
$q$ sufficiently small (large amount of resources) $p^\star(k)$ assumes the shape (\ref{eq:core}), whereas for 
large $q$ (small amount of resources) the width of the peak at $k=0$ increases and the second peak moves to $K\gg K_c$.

\subsection{Values of process integrity measures before and after attacks}

In Figure \ref{fig:Gamma_species} we plot the integrity measure $\Gamma[p,q]$, for the 
different network distributions considered, before and after 
an optimal intelligent attack. We also show the values 
\clearpage

\noindent
that the integrity 
measure would take after a random attack (where sites are picked up at random and removed 
until resources are exhausted), with the same attack resource

 limit. 
The node attack cost function and resource limit chosen are 
$\phi(k)=k^{\zeta}(k-1)$ and $\kappa^{-1}=\bra\phi(k)\ket_{\mathcal P}/q$, respectively, 
with $\zeta=0,1,2$ and $q>1$.
In addition we show the lower and upper 
bounds  (\ref{eq:GammaBoundsI}) on the process integrity measure after an optimal intelligent attack $q^\star[p]$, 
as described by the protocol in the previous section,
 as dashed and dotted lines. 
As expected, 
the data points for random attacks always coincide with the dotted line of the theoretical 
upper bound (since such strategies formed the basis from which the upper bound was derived).
Optimally attack resistant degree distributions are expected and indeed seen to be the ones 
for which the integrity measure of the network process 
after optimal intelligent attack is highest, 
but when quantified via $\Gamma[p,q]$ as in the figure one is struck by how well they perform, i.e. 
by the remarkably small reduction in the process integrity measure which they 
exhibit. 
For $\zeta=1$, the theoretical 
lower and upper bounds coincide with each other, 
and with the integrity measure values 
for random attacks and and optimal intelligent attacks.
Here $\phi(k)=k(k-1)$, so costs and benefit for the attacker of degree knowledge 
balance each other out,  and we have already showed that 
there is then no scope for the intelligent attacker to improve on the damage 
inflicted by random attacks. One can often understand the actual values 
obtained for $\Gamma[p,q]$. 
The relative reduction
of the integrity measure before and after  random attacks,  for instance, 
can be calculated from (\ref{eq:GammaBoundsI}), and for our 
resource limit $\kappa^{-1}=\bra \phi(k)\ket_{\mathcal P}/q$,
this gives $(\Gamma_{\rm before}-\Gamma_{\rm after})/\Gamma_{\rm before}=
\bra\phi(k)\ket_{\mathcal P}/q\bra\phi(k)\ket$. 
For $\zeta=0$ this is always equal to $1/q$;
for $\zeta=1, 2$ it is small for degree distributions with 
large second and third moments; much larger than for a Poissonian distribution. 
Similarly, for optimal intelligent attacks equation (\ref{eq:GammaBoundsI}) yield an upper bound 
on the relative attack-induced reduction of the process integrity measure:   
$(\Gamma_{\rm before}-\Gamma_{\rm after})/\Gamma_{\rm before}
\leq
\big[\bra k^{\zeta}(k-1)\ket_{\mathcal P}/q\bra k(k-1)\ket\big]{\rm max}_{k\in S}\{k^{1-\zeta}\}$. 
For $\zeta=1,2$, this change is again small  for optimally resistant and biological networks, as a result of their large degree variance
(except for C. Jejuni, which is distinct due to an unusually large average connectivity).

Our results 
re-confirm that random and hub-targeted attacks 
have similar effects on Poissonian graphs, as often remarked in literature, 
due to the large homogeneity of the degrees. For regular graphs they would have produced 
identical results. However, one should be careful in concluding from this that processes running on 
Poissonian networks are hence the most resistant ones against hub removal. On the contrary, 
Fig. \ref{fig:Gamma_species} shows that they are the most vulnerable ones, as their post-attack 
integrity measure is smallest. 
Interestingly, we find that processes running on networks produced by a preferential attachment mechanism 
are more resistant than those running on Poissonian networks, against both 
random attacks {\it and} optimal intelligent attacks. All this is due to the profound impact of 
resource constraints on the network resilience problem. 
Moreover, the degree distributions found in biological PINs generally exhibit, in turn, higher values for the post-attack 
process integrity measure than both Poissonian and preferential attachment networks, for random attacks 
and optimal intelligent attacks.
A final feature emerging from Fig. \ref{fig:Gamma_species} 
is that, while overall more robust compared to their preferential attachment and Poisonnian counterparts, 
biological networks seem significantly more resilient against random attacks than
against hub-targeted attacks (see the top two panels with $\zeta=0$, where the optimal attack indeed targets hubs).

\subsection{Connection with results of previous studies - fraction of removed nodes}

\begin{figure}[t]
\hspace*{5mm}
\begin{picture}(0,220)
\put(102,200){\includegraphics[width=29\unitlength,angle=-90]{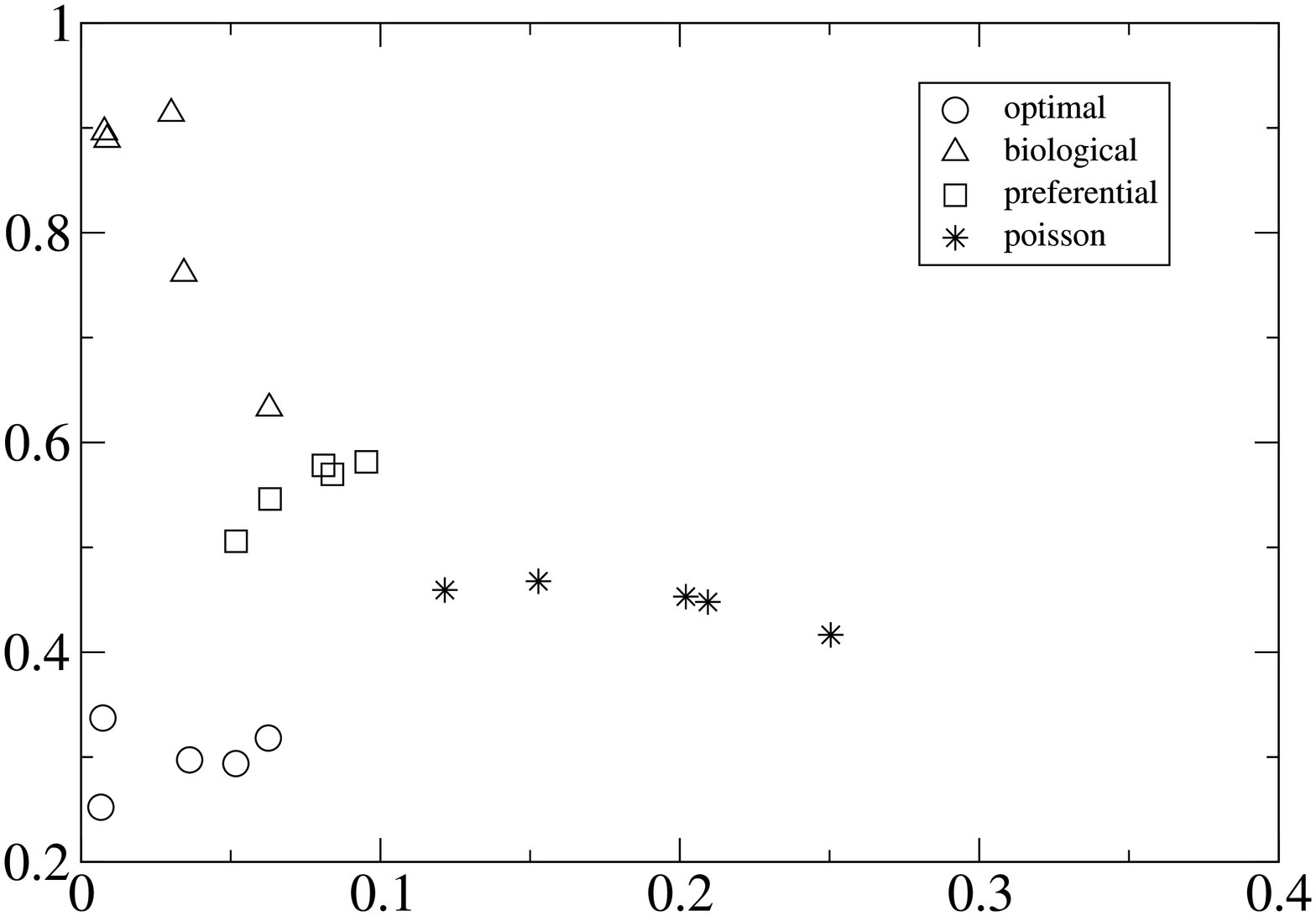}}
\put(-65,145){\Large $\frac{\Delta\Gamma}{\Gamma}$}
\put(60,53){$f$}
\put(310,200){\includegraphics[width=35.5\unitlength,angle=-90]{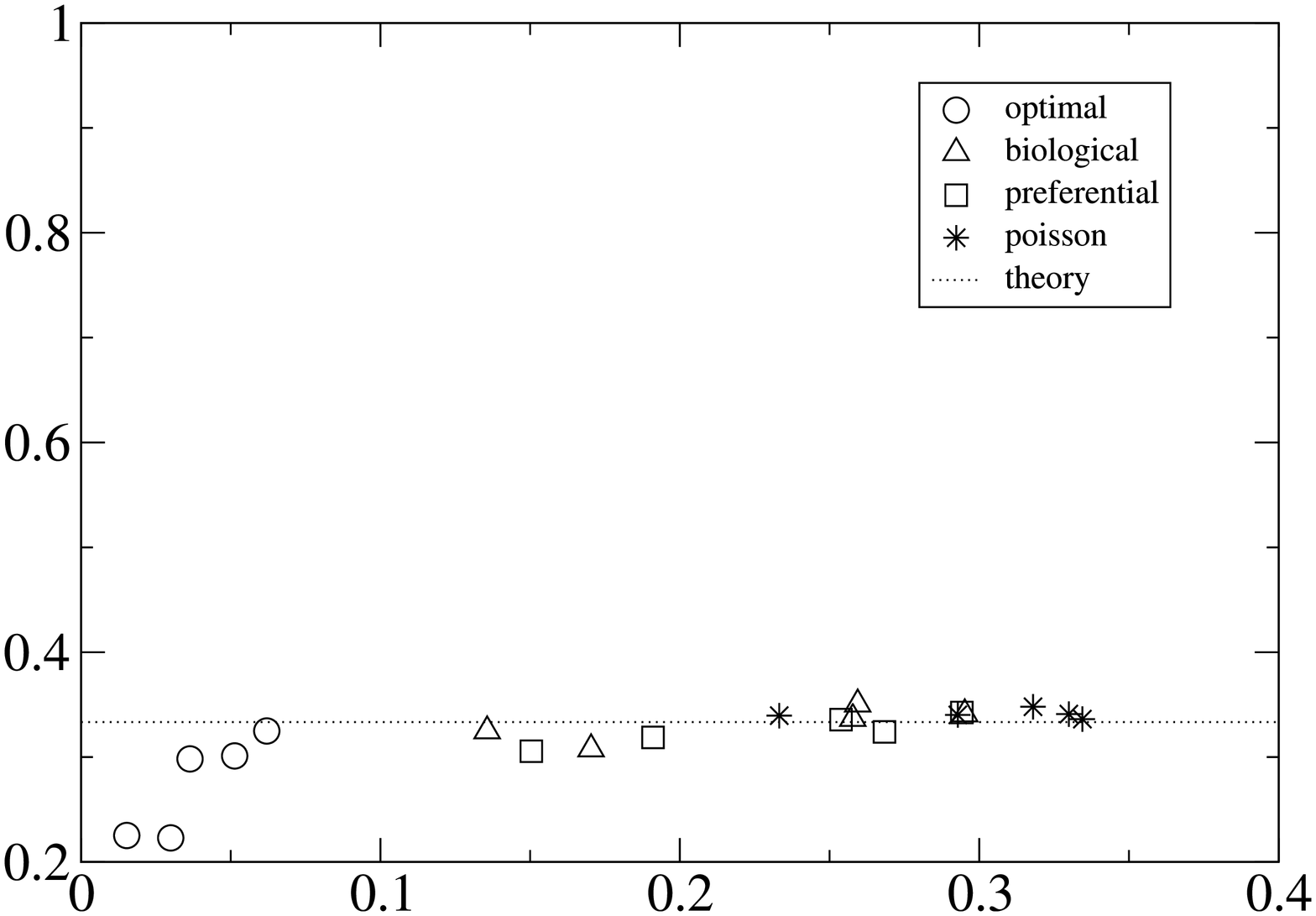}}
\put(270,53){$f$}
\end{picture}
\vspace*{-18mm}
\caption{Scatter plots of the relative variation $\Delta\Gamma/\Gamma$ of the integrity measure
versus the fraction $f$ of sites removed under optimal intelligent (left)
and random (right) attacks, 
with $\phi(k)=k-1$ and 
$\kappa^{-1}=\bra\phi(k)\ket/3$. Different markers correspond to 
 different  network families. 
The biological family is composed of the five experimentally determined PINs of 
Table \ref{tab:properties}. 
The other families are the synthetically generated counterparts of the biological PINs, 
with the same size and average connectivity,  
but different degree distributions (Poissonian, preferential attachment and optimally 
resistant against intelligent attack). 
}
\label{fig:removed}
\end{figure}

Previous studies of network resilience, based on analysis of static topological properties of networks under attacks,
had shown that power law networks (such as the ones produced by a preferential attachment mechanism)
are more resistant than Poissonian ones against random removal of 
a {\it fixed fraction} of nodes (see e.g. \cite{DorMen00,Bar02}), but are very vulnerable against hub removal 
\cite{AlbJeoBar00,Bar03}. In the light of our new results, one may wonder how the fraction $f$ of nodes removed 
varies among different degree distributions, when considering attacks constrained by degree dependent node removal costs, 
with limited attack resources. The results of numerical explorations  
for $\zeta=0$ (where optimal attacks will target hubs) and $q=3$ are shown in Fig. \ref{fig:removed}, in the form of  
scatter plots of the relative variation $\Delta \Gamma/\Gamma=(\Gamma_{\rm before}-\Gamma_{\rm after})/\Gamma_{\rm before}$ 
of the integrity measure under optimal hub-targeted
attacks  versus the fraction of sites removed (left), and under 
random attacks  versus the fraction of nodes removed (right).
The dotted line in the latter plot shows the theoretically predicted (constant) value of the relative variation of the process integrity measure
under random attacks for $\zeta=0$.
Figure \ref{fig:removed} 
reveals  that biological PINs are indeed affected by a dramatic 
drop in the integrity measure under hub removal, even for tiny fractions of removed nodes; this confirms our intuition that 
their observed resilience depends crucially on having degree statistics such that attack costs prevent intelligent attackers from removing significant numbers of hubs.  
The same statement is expected to apply to any random graph 
drawn from the ensemble (\ref{eq:connectivity}), where the imposed 
local degrees are those of the biological networks.  
Finally, Fig.~\ref{fig:removed} also shows that, as expected, the fraction of sites removed during hub-targeted attacks in a Poissonian 
graph, with fixed attack resources and when the node removal cost function is monotonically 
increasing with $k$, is considerably larger than in power law graphs. We conclude that the often claimed 
superiority of Poissonian networks over power law graphs for hub targeted attacks is strictly a consequence 
of the decision to keep the fraction of removed sites fixed. 
This is consistent with the findings in \cite{MagLatGui09}, where it was argued 
that power law networks are no longer more fragile than Poissonian graphs against hub targeted attacks when one 
looks at the number of removed {\it links}, and that the efficiency of hub removal in power law graphs 
would mainly lie  in the fact that this removes many more links than it would have in Poissonian graphs.

\begin{figure}[t]
\hspace*{-28mm}
\begin{picture}(200,199)
\put(200,180){\includegraphics[width=36\unitlength,angle=-90]{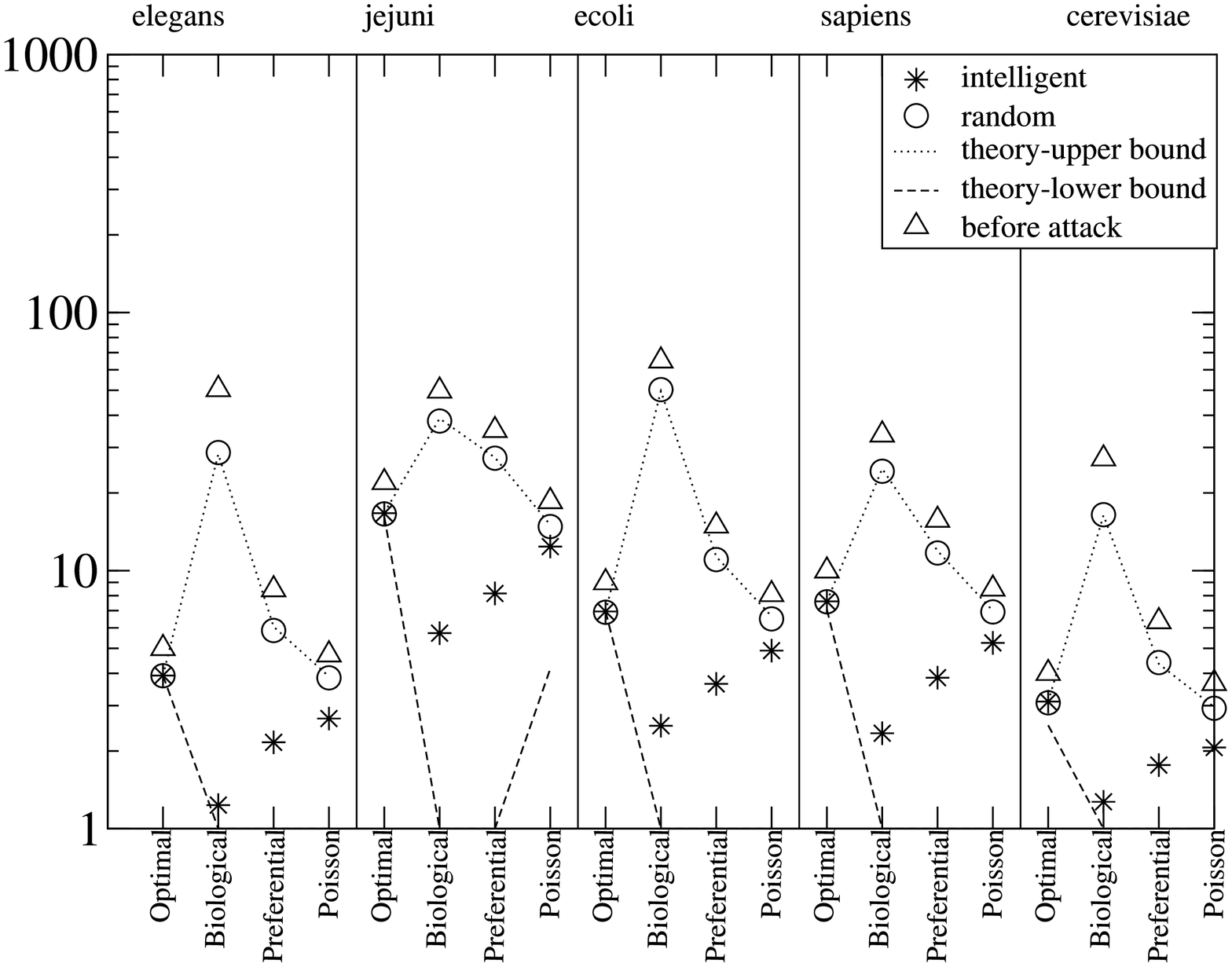}}
\put(29,100){\rotatebox{90}{$\Gamma[p,q]+1$}}
\put(295,180){\includegraphics[width=31\unitlength,angle=-90]{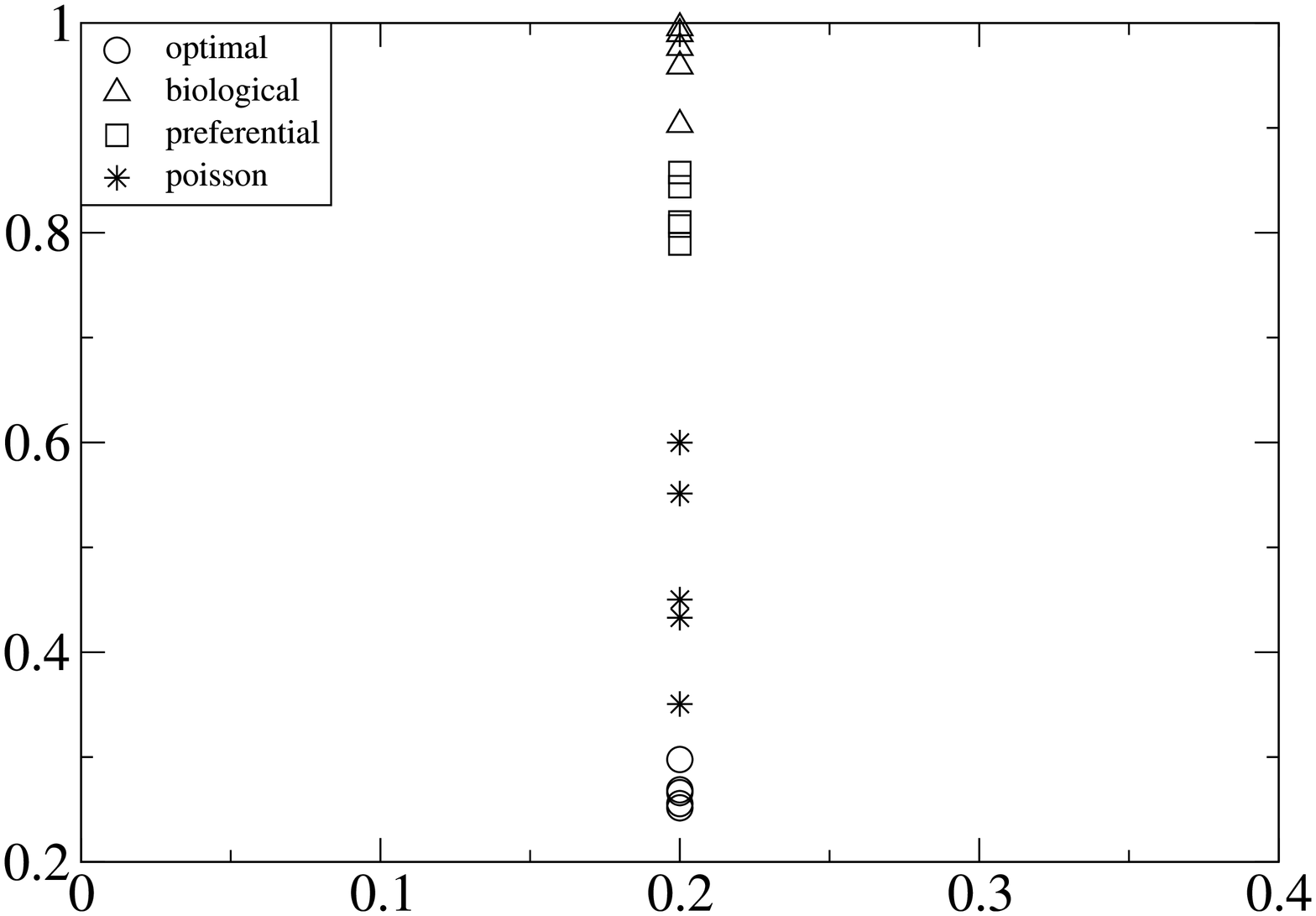}}
\put(270,100){\rotatebox{90}{$\Delta\Gamma/\Gamma$}}
\put(390,13){$f$}
\end{picture}
\vspace*{-2mm}
\caption{Left: process integrity measure $\Gamma[p,q]$ before and after optimal intelligent and random attacks, 
when the node removal cost function is a constant, $\phi(k)=\kav-1$, and attack resources are constrained according to
$\kappa^{-1}=\bra\phi(k)\ket/5$.
The theoretical upper and lower bounds 
of (\ref{eq:GammaBoundsI}) are shown  as dotted and dashed lines, respectively.
Right: Scatter plot of the relative variation $\Delta\Gamma/\Gamma$ of the integrity measure
versus the fraction of sites removed under intelligent attacks,  
with attack resources constrained according to $\kappa^{-1}=\bra\phi(k)\ket/5$ and with 
node attack cost function $\phi(k)=\kav-1$. Here different markers correspond to 
 different  network families, similar to Fig. \ref{fig:removed}.}
\label{fig:Gamma_species_const}
\end{figure}

In order to make contact with earlier results in literature, we consider 
below optimal attacks calculated for the constant cost function $\phi(k)=\phi$. Here 
the effects of attack costs should vanish from the problem, and our attacks should reduce to those where
the fraction $f$ of degrees to be removed is kept fixed.
In fact, from the resource constraint one has 
$f=\sum_k p(k)q(0|k)=1/\kappa \phi$.
We plot in Fig. \ref{fig:Gamma_species_const} (left) the integrity measure $\Gamma[p,q]$  for the different 
networks considered so far, before and after optimal intelligent and random attacks, 
with constant cost function $\phi=\kav -1$ and resources 
$\kappa^{-1}=\phi/5$. In the right panel we show a scatter plot of 
the relative variation of the process integrity measure under 
optimal intelligent attack 
versus the fraction of removed sites, similar to Fig.~\ref{fig:removed}.
The fraction of removed sites is now constant, as expected,  
and indeed equals $1/5$ for our choice of the resource limit.
We see that for the constant node removal cost function the dependence of the post-attack integrity measure on the degree distribution 
is drastically different from that in the case of monotonically increasing cost functions, 
and we retrieve the old 
results known from literature: power law networks are now more resilient 
than Poissonian ones against random attacks, but are extremely sensitive to hub removal.

\subsection{Misinformation}

\begin{figure}[t]
\hspace*{-6mm}
\begin{picture}(0,220)
\put(320,80){\includegraphics[width=30\unitlength,angle=-90]{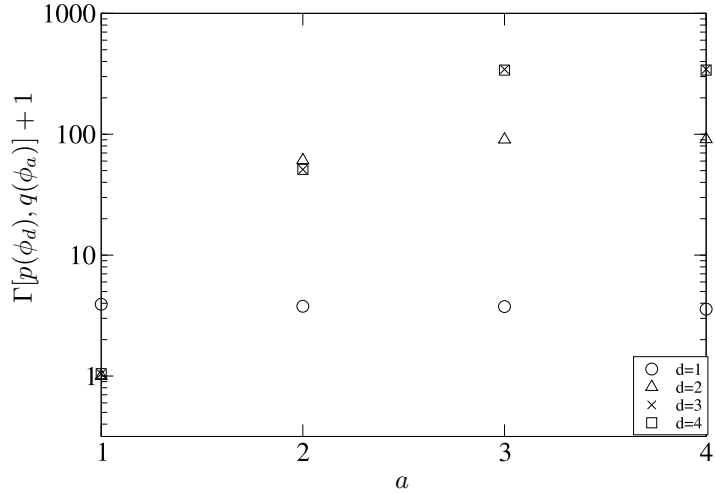}}
\put(85,100){\rotatebox{90}{$\Gamma[p(\phi_d),q(\phi_a)]+1$}}
\put(230,30){$a$}
\end{picture}
\vspace*{-9mm}
\caption{Log-plot of the post-attack integrity measure for the case where assumed and actual node removal costs and resource constraints 
need not be identical. 
Our network has  the size and average connectivity of the C. Elegans PIN. The node removal cost function is taken from a family 
$\phi_\ell(k)$ and the resource constraint is $\kappa_\ell^{-1}=\bra\phi_\ell(k)\ket/q_\ell$, with $\ell\in\{1,2,3,4\}$
(see main text for further details). 
The defender assumes that $\ell=d$ and chooses the associated optimally resistant degree distribution, whereas the actual 
value is $\ell=a$, and the attacker bases his strategy on the latter. The defender is optimally prepared only for $a=d$.
}
\label{fig:misinformation}
\end{figure}

We finally illustrate briefly the possible effects on the network resilience problem of misinformation, 
i.e. a situation where  a network is designed to be optimally 
resistant against an optimal intelligent attack on the basis of a node removal cost function
$\phi_d(k)$ and a resource limit $\kappa_d^{-1}$,  but where in fact it faces an optimal intelligent attack constrained by 
an actual cost  function $\phi_a(k)$ and with resource limit $\kappa_a^{-1}$. 
Here the cost functions $\phi_\ell(k)$  and resource limits $\kappa^{-1}_{\ell}$ are defined as follows (with integer $\ell\geq 1$):
$\phi_1(k)=\kav -1$, $\phi_{\ell>1}(k)=k^{\ell-2}(k-1)$, and $\kappa_\ell^{-1}=\bra \phi_\ell(k)\ket/q_\ell $ with $q_1=5$ and $q_{\ell>1}=3$.
Note that for $a=3$ the intelligent attacks in fact reduce to random ones. The results of our numerical explorations 
are shown in Fig \ref{fig:misinformation}.
For every attack, the distribution for which the post-attack process integrity measure is largest 
is indeed seen to be the one which is optimally resistant to the actual  attack, i.e. the choice $d=a$ 
(for $d=3,4$ the optimally resistant degree distributions are very similar, and their 
behaviour is almost identical). 
Fig~\ref{fig:misinformation} suggests that, as long as the node removal cost functions are monotonically increasing with the node degree (i.e. for $a\geq 2$),
networks which are optimally resistant against non-hub attacks ($a=3,4$) are reasonably 
resistant against hub-targeted attacks ($a=2$), whereas networks which are well prepared against hub-targeted attacks 
behave quite poorly when subjected to non-hub attacks ($a=3,4$). 
In other words, degree statistics designed to be optimally resistant against  
hub removal appear to be quite sensitive to misinformation, whereas those optimally resistant to 
random attacks suffer less from misinformation, at least as long as the node removal cost function is monotonically increasing with the node degree. 

\section{Discussion}

Many research papers have been devoted recently to the resilience of networks 
under attacks. Most study resilience in terms of the behaviour of static 
properties of networks under random and inteligent removal of a 
{\em fixed fraction} of sites or bonds.
Results obtained empirically \cite{AlbJeoBar00,BroKumMagRagRajStaTomWie00}
or analytically (within mean-field and asymptotic approximations)  
\cite{CohEreBenHav00,CohEreBenHav01,CohHavBen03,CalNewStrWat00,Newman03}
have shown that
power law networks are more resistant than Poissonian ones against 
random attacks (see e.g. \cite{DorMen00,Bar02}), but 
are very vulnerable against hub removal \cite{AlbJeoBar00,Bar03}. 
In contrast, more recent studies \cite{MagLatGui09} suggest that power law networks are not 
more fragile than Poissonian graphs against hub-targeted attacks when one 
looks at the number of removed {\it links} (as opposed to nodes).

In this paper we have sought to study network resilience in a more realistic setting, 
where attackers have {\it fixed resources} and where removing or disrupting 
a node carries a cost for the attacker which 
depends on the degree of the disrupted node. We quantify the 
resilience of the system in terms of the process for which the network acts as infrastructure,
based on determining the critical temperatures for the onset of various types of global order that could be envisaged
(the resulting network integrity measure is only weakly dependent upon the specific choices made). 
This formulation also allows for attacks involving partial disruption of individual nodes, which would have been inaccessible to the techniques 
normally used when studying network resilience, such as percolation theory.
 We can define precisely the most damaging attack strategy, given knowledge of the degree sequence of a network, and  
for any given node removal cost function. In addition we could 
subsequently define the optimal network topology, i.e. the degree distribution 
for which the integrity of the collective process is preserved best when attacked by 
a foe who employs the most damaging attack strategy. 

A network's  resilience against attacks is extremely sensitive 
to the dependence of the node removal cost function on the degree of the targeted node.
This dependence determines the crucial outcome of the competition in such scenarios between the benefit and the cost 
of attacking high-degree nodes. 
If we choose a trivial constant 
cost function, we retrieve results from literature on network resilience under 
random and targeted removal of fixed fractions of sites or bonds. 
However, as soon as one chooses more realistic
node removal cost functions, that increase sufficiently fast with the node's degree, 
power law networks are found to be more resistant  
than Poissonian ones, even against optimized intelligent attacks.
Our results show that ``modular'' configurations 
with a core of nodes highly connected to each other, in a sea of (almost) disconnected 
nodes, and strongly disassortative configurations, are, depending on the attacker's resources,  
the most resistant ones against hub-targeted attacks, respectively. Broad distributions with 
fat tails are the best defence against random and low degree targeted attacks. 
We also touched briefly upon the effects of misinformation, where a network is designed to be 
optimally resistant to a certain attack, whereas it actually faces a different one.
Results suggest that for monotonically increasing cost functions, degree distributions 
with fat tails are much less sensitive to misinformation effects.

Upon comparing real protein interaction networks with random networks of the 
same size and average degree, we found that the attack resilience of the biological networks is superior to that of 
 power law and Poissonian ones, even against optimized intelligent attacks. 
It may be that topological   properties beyond the degree sequence play an important
role here, and this deserves further investigation.  
In particular, one could calculate 
the integrity measure for processes supported by networks drawn from ensembles 
tailored to the production of graphs with built-in structure beyond that imposed by the degree distribution, 
along the lines of
\cite{VicCoo08,AnnCooFerKleFra09}. Another direction for future work 
may be to consider graph ensembles in which both the network topologies and 
the node removal cost 
functions involve hidden variables. 

Our paper emphasises the importance of distinguishing between different classes of network attacks 
on the basis of the node removal cost function and resource limitations imposed upon the attacker,   
and of studying and quantifying network resilience strictly within a given class of attacks. 
Previously proposed conclusions about the vulnerability of power law 
networks against intelligent attacks should be moderated in all cases 
where there is no compelling reason  to assume 
that the cost to attackers of node removal is independent of the node degrees.

\subsection*{Acknowledgements}

One of the authors (ACCC) would like to thank the Engineering and Physical Sciences Research Council (UK) for support
in the form of a Springboard Fellowship.

\section*{References}

\bibliography{net_references}
\bibliographystyle{unsrt}

\appendix

\section{Equilibrium analysis for model A}
\label{app:spins}

Most of the derivations in both appendices follow the lines of similar calculations in e.g. \cite{VicCoo08,
CooSkaPerPerHatWemNik05}, and we will hence be brief and highlight only crucial steps to indicate the changes
generated by the introduction of the attack variables $\{\xi_i\}$. Following \cite{AnnCooFerKleFra09} we use the property 
that with $\bra k\ket=N^{-1}\sum_i k_i$ the ensemble (\ref{eq:connectivity}) is identical to 
\begin{eqnarray}
{\rm Prob}(\bc)&=& \frac{\delta_{\bk,\bk(\bc)}}{{\cal Z}} \prod_{i<j}\Big[\frac{\bra k\ket}{N}\delta_{c_{ij},1}
\!+\!(1\!-\!\frac{\bra k\ket}{N})\delta_{c_{ij},0}\Big]
\label{eq:newer_connectivity}
\\
{\cal Z}&=&\sum_{\bc}\delta_{\bk,\bk(\bc)} \prod_{i<j}\Big[\frac{\bra k\ket}{N}\delta_{c_{ij},1}
\!+\!(1\!-\!\frac{\bra k\ket}{N})\delta_{c_{ij},0}\Big] 
 \end{eqnarray}

\subsection{Derivation of saddle-point equations}

We write the  Kronecker $\delta$s of the degree constraints in integral form, and we introduce the short-hands $\bsigma_i=(\sigma_i^1,\ldots,\sigma_i^n)\in\{-1,1\}^n$
 so that
\begin{eqnarray}
\hspace*{-15mm}
\overline{f}_{\!A}&=&
\lim_{N\to\infty}\lim_{n\to 0}\frac{1}{\beta nN}\Big\{
\log {\mathcal Z}-\log
\sum_{\bsigma_1\ldots\bsigma_N}\int_{-\pi}^\pi\prod_i\Big[\frac{\rmd\omega_i}{2\pi}\rme^{\rmi\omega_i k_i}\Big]
\nonumber
\\
\hspace*{-15mm}
&&\hspace*{15mm} \times
\prod_{i<j}\Big(
1
+\frac{\bra k\ket}{N}[\int\!\rmd J~P(J)\rme^{\beta J\xi_i\xi_j\bsigma_i\cdot\bsigma_j-\rmi(\omega_i+\omega_j)}\!
-1]\Big)\Big\}
\nonumber
\\
\hspace*{-15mm}
&=&
\lim_{N\to\infty}\lim_{n\to 0}\frac{1}{\beta nN}\Big\{
\log {\mathcal Z}-\log
\sum_{\bsigma_1\ldots\bsigma_N}\int_{-\pi}^\pi\prod_i\Big[\frac{\rmd\omega_i}{2\pi}\rme^{\rmi\omega_i k_i}\Big]
\nonumber
\\
\hspace*{-15mm}
&&\times\exp\Big[
\frac{\bra k\ket}{2N}\sum_{ij} [\int\!\rmd J~P(J)\rme^{\beta J\xi_i\xi_j\bsigma_i\cdot\bsigma_j-\rmi(\omega_i+\omega_j)}\!
-1]+\order(N^{0})\Big]\Big\}
\end{eqnarray}
We proceed by introducing for $\bsigma\in\{-1,1\}^n$ and $\xi\in\Xi$
the functions $D(\xi,\bsigma|\{\bsigma_i,\omega_i,\epsilon_i\})=
N^{-1}\sum_i\delta_{\xi,\xi_i}\delta_{\bsigma,\bsigma_i}\rme^{-\rmi\omega_i}$. They are introduced via the substitution of integrals
over appropriate $\delta$-distributions, written in integral form:
\begin{eqnarray}
1&=& \int\!\frac{\rmd D(\xi,\bsigma)\rmd \hat{D}(\xi,\bsigma)}{2\pi/N}\rme^{\rmi N\hat{D}(\xi,\bsigma)[D(\xi,\bsigma)-
D(\xi,\bsigma|\{\bsigma_i,\omega_i,\epsilon_i\})]}
\end{eqnarray}
Upon using the short hand
$\{\rmd D \rmd \hat{D}\}=\prod_{\epsilon,\bsigma}D(\epsilon,\bsigma)\rmd \hat{D}(\epsilon,\bsigma)$ we then obtain
\begin{eqnarray}
\hspace*{-15mm}
\overline{f}_{\!A}&=&
\lim_{N\to\infty}\lim_{n\to 0}\frac{1}{\beta nN}\Big\{
\!\log {\mathcal Z}\!-\!\log
\int\!\{\rmd D\rmd \hat{D}\}\rme^{\rmi N\sum_{\epsilon\bsigma}\hat{D}(\epsilon,\bsigma)D(\epsilon,\bsigma)-\frac{1}{2}N\bra k\ket
+\order(\log N)}
\!\!\!\!
\nonumber
\\
\hspace*{-15mm}
&&\times\exp\Big[
\frac{1}{2}\bra k\ket N \sum_{\epsilon\epsilon^\prime}\sum_{\bsigma\bsigma^\prime}D(\epsilon,\bsigma)D(\epsilon^\prime\!,\bsigma^\prime)
\int\!\rmd J~P(J)\rme^{\beta J\epsilon\epsilon^\prime\bsigma\cdot\bsigma^\prime}\!
\Big]
\nonumber
\\
\hspace*{-15mm}
&&
\times \exp\Big[ N\sum_{\xi k} p(\xi,k)\log
\sum_{\bsigma}\int_{-\pi}^\pi\!\frac{\rmd\omega}{2\pi}\rme^{\rmi\omega k-\rmi
\hat{D}(\xi,\bsigma)\rme^{-\rmi\omega} }
\Big]
\Big\}
\end{eqnarray}
We next define $z=\lim_{N\to\infty}N^{-1}\log {\mathcal Z}$ (anticipating this limit to exist),
which allows us to evaluate $\overline{f}$ by steepest descent:
\begin{eqnarray}
\overline{f}_{\!A}&=&\lim_{n\to 0} \frac{1}{n}{\rm extr}_{\{D,\hat{D}\}} f_{n,A}[\{D,\hat{D}\}]
\label{eq:f_steepest_descent}
\\
f_{n,A}[\ldots]&=&
-\frac{1}{\beta}\Big\{
\rmi \sum_{\xi\bsigma}\hat{D}(\xi,\bsigma)D(\xi,\bsigma)-\frac{1}{2}\bra k\ket
-z
\nonumber
\\
&&
+\frac{1}{2}\bra k\ket \sum_{\xi\xi^\prime}\sum_{\bsigma\bsigma^\prime}D(\xi,\bsigma)D(\xi^\prime\!,\bsigma^\prime)
\int\!\rmd J~P(J)\rme^{\beta J\xi\xi^\prime\bsigma\cdot\bsigma^\prime}\!
\nonumber
\\
&&
 +\sum_{\xi k} p(\xi,k)\log
\sum_{\bsigma}\int_{-\pi}^\pi\!\frac{\rmd\omega}{2\pi}\rme^{\rmi\omega k-\rmi
\hat{D}(\xi,\bsigma)\rme^{-\rmi\omega} }
\Big\}
\label{eq:fD}
\end{eqnarray}
Extremization (\ref{eq:fD}) with respect to $\{D,\hat{D}\}$ gives the saddle-point equations:
\begin{eqnarray}
\hat{D}(\xi,\bsigma)&=&
\rmi\bra k\ket \sum_{\xi^\prime}\sum_{\bsigma^\prime}D(\xi^\prime\!,\bsigma^\prime)
\int\!\rmd J~P(J)\rme^{\beta J\xi\xi^\prime\bsigma\cdot\bsigma^\prime}
\\
 D(\xi,\bsigma)
 &=&\sum_k p(\xi,k)
 \frac{\int_{-\pi}^\pi\!\rmd\omega~\rme^{\rmi\omega (k-1)-\rmi
\hat{D}(\xi,\bsigma)\rme^{-\rmi\omega} } }
{\sum_{\bsigma^\prime}\int_{-\pi}^\pi\!\rmd\omega~\rme^{\rmi\omega k-\rmi
\hat{D}(\xi,\bsigma^\prime)\rme^{-\rmi\omega} }}
\end{eqnarray}
The second of these equations is simplified using the identity
\begin{eqnarray}
\int_{-\pi}^\pi\!\rmd\omega~\rme^{\rmi\omega\ell-\rmi
\hat{D}(\xi,\bsigma)\rme^{-\rmi\omega} } &=& \left\{\begin{array}{lll}
2\pi[-\rmi\hat{D}(\xi,\bsigma)]^{\ell}/\ell! &~~{\rm if}~~& \ell\geq 0
\\
0 &~~{\rm if}~~& \ell<0
\end{array}\right.
\label{eq:identity}
\end{eqnarray}
So, if we also re-define $\hat{D}(\xi,\bsigma)=\rmi \bra k\ket F(\xi,\bsigma)$, we arrive at
\begin{eqnarray}
F(\xi,\bsigma)&=& \sum_{\xi^\prime}\sum_{\bsigma^\prime}D(\xi^\prime\!,\bsigma^\prime)
\int\!\rmd J~P(J)\rme^{\beta J\xi\xi^\prime\bsigma\cdot\bsigma^\prime}
\label{eq:SP_F}
\\
 D(\xi,\bsigma)
 &=& \sum_{k>0}p(\xi,k)\frac{k}{\bra k\ket}
 \frac{F^{k-1}(\xi,\bsigma)}
{\sum_{\bsigma^\prime}F^k(\xi,\bsigma^\prime)}
\label{eq:SP_D}
\end{eqnarray}
We note that $\sum_{\xi}\sum_{\bsigma}D(\xi,\bsigma)F(\xi,\bsigma)=1$ at the saddle-point.
The term $z=\lim_{N\to\infty}N^{-1}\log {\mathcal Z}$ measures the
number of graphs in the ensemble. It follows from  $\lim_{\beta\to 0}(\beta \overline{f})=-\log 2$, giving
$z= \bra k\ket\log \bra k\ket-\bra k\ket -
\sum_{k} p(k)\log k!$,
and hence
\begin{eqnarray}
\overline{f}_{\! A}&=&
-\lim_{n\to 0}\frac{1}{\beta n}
\sum_{\xi k} p(\xi,k)\log \Big[ \sum_{\bsigma} F^k(\xi,\bsigma)\Big]
\label{eq:f_RSB}
\end{eqnarray}

\subsection{Replica symmetric theory}

To take the required limit $n\to 0$ in our formulae we make the replica-symmetric (RS) ansatz.
The order parameter $D(\xi,\bsigma)$ must now be invariant under all replica permutations, and thus have the following form:
\begin{eqnarray}
D(\xi,\bsigma)&=&\int\!\rmd h~ D(\xi,h)\frac{\rme^{\beta
h\sum_{\alpha}\sigma_\alpha}}{[2\cosh(\beta h)]^n}
\label{eq:RSansatz}
\end{eqnarray}
Via equations (\ref{eq:SP_F},\ref{eq:SP_D}) one then finds a similar structure  for $F(k,\bsigma)$,
\begin{eqnarray}
F(\xi,\bsigma)&=&  \int\!\rmd h~F(\xi,h)~ \rme^{\beta h\sum_\alpha \sigma_\alpha}
\label{eq:RSansatzF}
\end{eqnarray}
and in the limit $n\to 0$, after some standard manipulations, a closed set of transparant equations for the RS order parameters $D(\xi,h)$ and $F(\xi,h)$:
\begin{eqnarray}
\hspace*{-20mm}
F(\xi,h)&=& \sum_{\xi^\prime}\!\int\!\rmd h^\prime \rmd J~ D(\xi^\prime\!,h^\prime)P(J)
\delta\Big[h\!-\!\frac{1}{\beta}\atanh[\tanh(\beta J\xi\xi^\prime)\tanh(\beta h^\prime)]\Big]
\label{eq:RS1}
\\
\hspace*{-20mm}
 D(\xi,h)
 &=&
 \sum_k p(\xi,k)
\frac{k}{\bra k\ket}
 \frac{\int\!\prod_{\ell< k}[\rmd h_\ell F(\xi,h_\ell)] \delta[h-\sum_{\ell< k}h_\ell]}
{[\int\!\rmd h^\prime F(\xi,h^\prime)]^k }
\label{eq:RS2}
\end{eqnarray}
We note upon integrating and combining these equations that $\int\!\rmd h~F(\xi,h)=\sum_{\xi^\prime}\int\!\rmd h~D(\xi^\prime,h)=1$. 
This enables us to write $F(\xi,h)=F(h|\xi)$ with $\int\!\rmd h~F(h|\xi)=1$,
which gives immediate probabilistic interpretations of the functions $F(h|\xi)$.
Upon eliminating $D(\xi,h)$ the RS saddle-point equations then take the new form
\begin{eqnarray}
F(h|\xi)&=& \sum_{k \xi^\prime}
 p(\xi^\prime\!,k)
\frac{k}{\bra k\ket}
 \int\!\rmd J~P(J)\int\!\prod_{\ell< k}[\rmd h_\ell F(h_\ell|\xi^\prime)]
 \nonumber
 \\
 &&\times~
\delta\Big[h-\frac{1}{\beta}\atanh[\tanh(\beta J\xi\xi^\prime)\tanh(\beta \sum_{\ell< k}h_\ell)]\Big]
\label{eq:RS}
\end{eqnarray}
Clearly $F(h|0)=\delta(h)$.
To identify the relevant observables and calculate for $\bsigma\in\{-1,1\}^n$ the quantity $P(\xi,k,\bsigma)=\lim_{N\to\infty} N^{-1}\sum_i\overline{\bra \delta_{\xi,\xi_i}\delta_{k,k_i}\delta_{\bsigma,\bsigma_i}\ket}$ one uses the alternative form of the replica identity, viz.
\begin{eqnarray}
\hspace*{-10mm}
\overline{\bra g(\bsigma)\ket}&=&
\overline{\left[\frac{\sum_{\bsigma}g(\bsigma)e^{-\beta H(\bsigma)}}{\sum_{\bsigma}e^{-\beta H(\bsigma)}}\right]}=
\lim_{n\to 0}\sum_{\bsigma^1\ldots\bsigma^n} \overline{g(\sigma^1)e^{-\beta \sum_{\alpha=1}^n H(\bsigma^\alpha)}}
\end{eqnarray}
Upon also making the RS ansatz this results in
\begin{eqnarray}
P_{\rm RS}(\xi,k,\bsigma)&=& p(\xi,k)\int\!\rmd h~W(h|\xi,k)\frac{\rme^{\beta h \sum_{\alpha}\sigma_\alpha}}{[2\cosh(\beta h)]^n}
\label{eq:spinstats}
\\
W(h|\xi,k)&=& \int\!\prod_{\ell\leq k}[\rmd h_\ell F(h_\ell|\xi)]\delta[h-\sum_{\ell\leq k}h_\ell]
\label{eq:effective_fields}
\end{eqnarray}
The measure $W(h|\xi,k)$ is the effective field distribution for those sites where $(\xi_i,k_i)=(\xi,k)$.
We note that $W(h|0,k)=\delta(h)$. With $W(h)=\sum_{\xi k}p(\xi,k) W(h|\xi,k)$ we can write the conventional scalar order parameters $m=\lim_{N\to\infty}N^{-1}\sum_i\overline{\bra\sigma_i\ket}$ and $q=\lim_{N\to\infty}N^{-1}\sum_i\overline{\bra\sigma_i\ket^2}$  in their familiar forms
\begin{eqnarray}
m=\int\!\rmd h~W(h)\tanh(\beta h),~~~~~~~~q= \int\!\rmd h~W(h)\tanh^2(\beta h)
\end{eqnarray}
The subset of sites with $(\xi_i,k_i)=(\xi,k)$ can be regarded
as sublattices in the sense of \cite{WemCoo03}, and we can define sublattice magnetizations $m(\xi,k)$ via
$m(\xi,k)=\int\!\rmd h~W(h|\xi,k)\tanh(\beta h)$, such that $m=\sum_{\xi k}p(\xi,k)m(\xi,k)$.
In the limit $T\to\infty$ (i.e. $\beta\to 0$) the only solution of (\ref{eq:RS})
is as always the trivial paramagnetic (P) one: $F(h|\xi)=\delta(h)$. This is a saddle-point at any temperature, but can become unstable
in favour of ferromagnetic (F) or spin-glass (SG) states as $T$ is lowered.

\subsection{Continuous phase transitions away from the paramagnetic state}

Continuous bifurcations away from the trivial state are found as usual by expanding (\ref{eq:RS})
in moments of $F(h|\xi)$, assuming
the existence of a small parameter $\epsilon$ with  $0\!<\!|\epsilon|\!\ll\! 1$ such that $\int\!\rmd h~h^\ell F(h|\xi)=\order(\epsilon^\ell)$.
With some foresight we define a function $\gamma(\xi)$ and two $|\Xi|\!\times\!|\Xi|$ matrices $M^{(\ell)}(\beta)$ with entries
$M^{(\ell)}_{\xi \xi^\prime}(\beta)$, for $\ell\in\{1,2\}$:
\begin{eqnarray}
\gamma(\xi)&=& \bra k\ket^{-1}\sum_k p(\xi,k)k(k-1)
\label{eq:gammaxi}
\\
M^{(\ell)}_{\xi\xi^\prime}(\beta)&=&\gamma(\xi^\prime)\int\!\rmd J~P(J)\tanh^\ell(\beta J\xi\xi^\prime)
\end{eqnarray}
We define $\lambda^{(\ell)}_{\rm max}(\beta)$ as the largest eigenvalue  of $M^{(\ell)}(\beta)$.
If the first order to bifurcate away from $F(h|\xi)=\delta(h)$ is $\epsilon^1$, the bifurcation is towards a state where $m\neq 0$, i.e. describing a P$\to$F transition. Upon multiplying both sides of (\ref{eq:RS}) by $h$ and integrating over $h$, the bifurcation condition for this is found to be
\begin{eqnarray}
{\rm P}\to{\rm F}:&~~~& \lambda^{(1)}_{\rm max}(\beta)=1
\label{eq:PtoF}
\end{eqnarray}
If instead the first order to bifurcate is $\epsilon^2$, the bifurcating new state has $m=0$ and $q>0$, describing a P$\to$SG transition. Upon multiplying both sides of (\ref{eq:RS}) by $h^2$ and integrating over $h$, the bifurcation condition for this is found to be
\begin{eqnarray}
{\rm P}\to{\rm SG}:&~~~& \lambda^{(2)}_{\rm max}(\beta)=1
\label{eq:PtoSG}
\end{eqnarray}
We focus on a specific simple bond distribution, the binary 
$P(J)=\frac{1}{2}(1\!+\!\eta)\delta(J\!-\!J_0)+\frac{1}{2}(1\!-\!\eta)\delta(J\!+\!J_0)$ (with $J_0\geq 0$), where the matrices
$M^{(\ell)}(\beta)$ take the simple form:
\begin{eqnarray}
&&
\hspace*{-15mm}
M^{(1)}_{\xi\xi^\prime}(\beta)=\eta\tanh(\beta J_0\xi\xi^\prime)\gamma(\xi^\prime)
~~~~~~~~
M^{(2)}_{\xi\xi^\prime}(\beta)= \tanh^2(\beta J_0\xi\xi^\prime)\gamma(\xi^\prime)
\label{eq:MmatricesA}
\end{eqnarray}

\section{Equilibrium analysis for model B}
\label{app:oscillators}

\subsection{Derivation of saddle-point equations}

The calculation for coupled oscillators is initially very similar to the previous one, with summations replaced by integrations. 
The main differences start at the introduction of the replica-symmetry ansatz; from then onwards we have to implement appropriate adaptations of the calculation for XY spins in 
\cite{CooSkaPerPerHatWemNik05}
(an alternative route would be to adapt the cavity-based analysis in \cite{SkaPerHat05}).
As before we write  degree constraints in integral form, and we introduce the short-hands $\btheta_i=(\theta_i^1,\ldots,\theta_i^n)\in[-\pi,\pi]^n$
 so that
\begin{eqnarray}
\hspace*{-15mm}
\overline{f}_{\!B}&=&
\lim_{N\to\infty}\lim_{n\to 0}\frac{1}{\beta nN}\Big\{
\log {\mathcal Z}-\log
\int_{-\pi}^\pi\!
\rmd\btheta_1\ldots \rmd \btheta_N
\int_{-\pi}^\pi\prod_i\Big[\frac{\rmd\omega_i}{2\pi}\rme^{\rmi\omega_i k_i}\Big] 
\nonumber
\\
\hspace*{-15mm}
&&
\hspace*{-7mm}\times \exp\Big[
\frac{\bra k\ket}{2N}\sum_{ij} [\int\!\!\rmd J~P(J)\rme^{\beta J\xi_i\xi_j\sum_\alpha\cos(\theta^\alpha_i-\theta^\alpha_j)-\rmi(\omega_i+\omega_j)}\!
-\!1]+\order(N^{0})\Big]\Big\}
\end{eqnarray}
We next introduce for $\btheta\in[-\pi,\pi]^n$ and $\xi\in\{0,1\}$
the functions $D(\xi,\btheta|\{\btheta_i,\omega_i,\epsilon_i\})=
N^{-1}\sum_i\delta_{\xi,\xi_i}\delta[\btheta,\btheta_i]\rme^{-\rmi\omega_i}$, via the substitution of functional integrals
over appropriate $\delta$-distributions, written in integral form.
With the short hand
$\{\rmd D \rmd \hat{D}\}=\prod_{\epsilon,\btheta}D(\epsilon,\btheta)\rmd \hat{D}(\epsilon,\btheta)$ we then obtain
an expression in the form of  path integral:
\begin{eqnarray}
\hspace*{-18mm}
\overline{f}_{\!B}&=&
\lim_{N\to\infty}\lim_{n\to 0}\frac{1}{\beta nN}\Big\{
\!\log {\mathcal Z}\!-\!\log\!
\int\!\{\rmd D\rmd \hat{D}\}\rme^{\rmi N\sum_{\epsilon}\int\!\rmd \btheta \hat{D}(\epsilon,\btheta)D(\epsilon,\btheta)-\frac{1}{2}N\bra k\ket
+\order(\log N)}
\!\!\!\!
\nonumber
\\
\hspace*{-18mm}
&&\times\exp\Big[
\frac{1}{2}\bra k\ket N \sum_{\epsilon\epsilon^\prime}\int\!\rmd \btheta\rmd \btheta^\prime D(\epsilon,\btheta)D(\epsilon^\prime\!,\btheta^\prime)
\!\int\!\rmd J~P(J)\rme^{\beta J\epsilon\epsilon^\prime\sum_\alpha \cos(\theta_\alpha-\theta^\prime_\alpha)}\!
\Big]
\nonumber
\\
\hspace*{-18mm}
&&
\times \exp\Big[ N\sum_{\xi k} p(\xi,k)\log
\int_{-\pi}^\pi\!\rmd \btheta\int_{-\pi}^\pi\!\frac{\rmd\omega}{2\pi}\rme^{\rmi\omega k-\rmi
\hat{D}(\xi,\btheta)\rme^{-\rmi\omega} }
\Big]
\Big\}
\end{eqnarray}
With $z=\lim_{N\to\infty}N^{-1}\log {\mathcal Z}_N=\bra k\ket\log \bra k\ket-\bra k\ket -
\sum_{k} p(k)\log k!$ (which has already been calculated earlier),
we evaluate $\overline{f}$ by steepest descent:
\begin{eqnarray}
\hspace*{-15mm}
\overline{f}_{\!B}&=&\lim_{n\to 0} \frac{1}{n}{\rm extr}_{\{D,\hat{D}\}} f_{n,B}[\{D,\hat{D}\}]
\label{eq:fB_steepest_descent}
\\
\hspace*{-15mm}
f_{n,B}[\ldots]&=&
-\frac{1}{\beta}\Big\{
\rmi \sum_{\xi}\int\!\rmd \btheta~\hat{D}(\xi,\btheta)D(\xi,\btheta)-\frac{1}{2}\bra k\ket
-z
\nonumber
\\
\hspace*{-15mm}
&&
+\frac{1}{2}\bra k\ket \sum_{\xi\xi^\prime}\int\!\rmd\btheta \rmd\btheta^\prime D(\xi,\btheta)D(\xi^\prime\!,\btheta^\prime)
\int\!\rmd J~P(J)\rme^{\beta J\xi\xi^\prime\sum_\alpha\cos(\theta_\alpha-\theta_\alpha^\prime)}\!
\nonumber
\\
\hspace*{-15mm}
&&
 +\sum_{\xi k} p(\xi,k)\log
\int_{-\pi}^\pi\!\rmd\btheta
\int_{-\pi}^\pi\!\frac{\rmd\omega}{2\pi}\rme^{\rmi\omega k-\rmi
\hat{D}(\xi,\btheta)\rme^{-\rmi\omega} }
\Big\}
\label{eq:fDB}
\end{eqnarray}
Functional variation of (\ref{eq:fB_steepest_descent}) with respect to $\{D,\hat{D}\}$, followed by application of (\ref{eq:identity}) and transformation via $\hat{D}(\xi,\theta)=\rmi \bra k\ket F(\xi,\btheta)$, gives the saddle-point equations
\begin{eqnarray}
F(\xi,\btheta)&=& \sum_{\xi^\prime}\int\!\rmd\btheta^\prime~D(\xi^\prime\!,\btheta^\prime)
\int\!\rmd J~P(J)\rme^{\beta J\xi\xi^\prime\sum_\alpha\cos(\theta_\alpha-\theta_\alpha^\prime)}
\label{eq:SP_FB}
\\
 D(\xi,\btheta)
 &=& \sum_{k>0}p(\xi,k)\frac{k}{\bra k\ket}
 \frac{F^{k-1}(\xi,\btheta)}
{\int\!\rmd\btheta^\prime~F^k(\xi,\btheta^\prime)}
\label{eq:SP_DB}
\end{eqnarray}
Again $\sum_{\xi}\int\!\rmd\btheta~D(\xi,\btheta)F(\xi,\btheta)=1$ at the saddle-point, and we obtain 
\begin{eqnarray}
\overline{f}_{\!B}&=&
-\lim_{n\to 0}\frac{1}{\beta n}
\sum_{\xi k} p(\xi,k)\log \Big[ \int\!\rmd \btheta~ F^k(\xi,\btheta)\Big]
\label{eq:f_RSBB}
\end{eqnarray}

\subsection{Replica symmetric theory}

For real-valued variables the replica-symmetric ansatz is less straightforward.
Permutation invariance with respect to $\btheta$ components now implies that $D(\xi,\btheta)$ and
$F(\xi,\btheta)$ are functional integrals over the space of normalized functions $P:[-\pi,\pi]\to \R$ (i.e. $\int_{-\pi}^\pi\!d\theta~P(\theta)=1$), 
with functional measures $W_D[\xi,\{P\}]$ and $W_F[\xi,\{P\}]$:
\begin{eqnarray}
D(\xi,\btheta)&=&\int\!\{\rmd P\}~W_D[\xi,\{P\}]\prod_\alpha P(\theta_\alpha)
\label{eq:RSansatzDB}
\\
F(\xi,\btheta)&=&\int\!\{\rmd P\}~W_F[\xi,\{P\}]\prod_\alpha P(\theta_\alpha)
\label{eq:RSansatzFB}
\end{eqnarray}
(we may use the same symbol $P$ as employed to define the bond probabilities via $P(J)$, the arguments will always prevent ambiguity).
Insertion of (\ref{eq:RSansatzDB},\ref{eq:RSansatzFB}) into the two equations (\ref{eq:SP_FB},\ref{eq:SP_DB}) then gives,
in the limit $n\to 0$ and after some manipulations, the following closed equations for the RS measures $W_D[\xi,\{P\}]$ and $W_F[\xi,\{P\}]$:
\begin{eqnarray}
\hspace*{-0mm}
W_F[\xi,\{P\}]&=& \sum_{\xi^\prime}\int\!\{\rmd P^\prime\} W_D[\xi^\prime\!,\{P^\prime\}]\int\!\rmd J P(J)
\nonumber
\\
&&\hspace*{5mm}\times 
\prod_\theta \delta\Big[P(\theta)\!-\!\frac{\int\!\rmd\theta^\prime \rme^{\beta J\xi\xi^\prime\cos(\theta-\theta^\prime)}P^\prime(\theta^\prime)}{2\pi I_0(\beta J\xi\xi^\prime)}\Big]
\\
\hspace*{-0mm}
W_D[\xi,\{P\}]&=&\sum_{k>0}\frac{p(\xi,k)k/\bra k\ket}{\Big[\int\!\{\rmd P^\prime\}W_F[\xi,\{P^\prime\}]\Big]^{k}}
\int\!\prod_{\ell<k}\Big[\{\rmd P_\ell\}W_F[\xi,\{P_\ell\}]\Big]\nonumber
\\
\hspace*{-0mm}&&\hspace*{5mm}
\times \prod_\theta \delta\Big[P(\theta)-\frac{\prod_{\ell<k}P_\ell(\theta)}{\int\!\rmd\theta^\prime
\prod_{\ell<k}P_\ell(\theta^\prime)}\Big]
\end{eqnarray}
Functional integration of both equations over $P$ shows that
$\int\{\rmd P\}W_F[\xi,\{P\}]=\sum_{\xi^\prime}\int\{\rmd P\} W_D[\xi^\prime,\{P\}]=1$.
This allows us to
write $W_F[\xi,\{P\}]=W_F[\{P\}|\xi]$ with $\int\{\rmd P\} W_F[\{P\}|\xi]=1$,
which allows also here for probabilistic interpretations of the order parameters $W_F[\xi,\{P\}]$, which are now functionals acting on the space of
probability distributions over the interval $[-\pi,\pi]$.
Upon eliminating $W_D[\xi,\{P\}]$ the RS saddle-point equations then take the following form (where $\theta,\theta^\prime\in[-\pi,\pi]$):
\begin{eqnarray}
W_F[\{P\}|\xi]&=& \sum_{k\xi^\prime}p(\xi^\prime,k)\frac{k}{\bra k\ket}\int\!\rmd J~ P(J)
\int\!\prod_{\ell<k}\Big[\{\rmd P_\ell\}W_F[\{P_\ell\}|\xi^\prime]\Big]\nonumber
\\
&&
\times
\prod_\theta \delta\Big[P(\theta)\!-\!\frac{\int\!\rmd\theta^\prime~ \rme^{\beta J\xi\xi^\prime\cos(\theta-\theta^\prime)}\prod_{\ell<k}P_\ell(\theta^\prime)}{2\pi I_0(\beta J\xi\xi^\prime)
\int\!\rmd\theta^\prime
\prod_{\ell<k}P_\ell(\theta^\prime)}\Big]
\label{eq:RSmodelB}
\end{eqnarray}
We observe that $W_F[\{P\}|0]=\prod_\theta \delta[P(\theta)-(2\pi)^{-1}]$.
To identify the physical meaning of our observables we define and calculate the quantity $P(\xi,k,\btheta)=\lim_{N\to\infty} N^{-1}\sum_i\overline{\bra \delta_{\xi,\xi_i}\delta_{k,k_i}\delta[\btheta,\btheta_i]\ket}$. Within the RS ansatz it is found to be
\begin{eqnarray}
\hspace*{-15mm}
P_{\rm RS}(\xi,k,\btheta)&=& p(\xi,k)\int\!\{\rmd P\}~W[\{P\}|\xi,k]\prod_\alpha P(\theta_\alpha)
\label{eq:spinstatsB}
\\
\hspace*{-15mm}
W[\{P\}|\xi,k]&=&\! \int\!\prod_{\ell\leq k}\Big[\{\rmd P_\ell\} W_F[\{P_\ell\}|\xi]\Big]\prod_\theta \delta\Big[P(\theta)\!-\!\frac{\prod_{\ell\leq k}P_\ell(\theta)}
{\int\!\rmd\theta^\prime\prod_{\ell\leq k}P_\ell(\theta^\prime)}\Big]
\label{eq:effective_fieldsB}
\end{eqnarray}
The functional measure $W[\{P\}|\xi,k]$ generalizes the concept of an effective field to an `effective' angle distribution of those oscillators
with $(\xi_i,k_i)=(\xi,k)$.
Note that $W[\{P\}|0,k]=\prod_\theta\delta[P(\theta)-(2\pi)^{-1}]$. With $W[\{P\}]=\sum_{\xi k}p(\xi,k) W[\{P\}|\xi,k]$ we can write the
conventional types of scalar order parameters
 in a compact form:
\begin{eqnarray}
\hspace*{-23mm}
\lim_{N\to\infty}\frac{1}{N}\!\sum_i\overline{\bra f(\theta_i)\ket}&=& \!\int\!\{\rmd P\}W[\{P\}]\int_{-\pi}^\pi\!\!\!\rmd\theta~P(\theta)f(\theta)
\\
\hspace*{-23mm}
\lim_{N\to\infty}\frac{1}{N}\!\sum_i\overline{\bra f(\theta_i)\ket\bra g(\theta_i)\ket}&=& \!
\int\!\{\rmd P\}W[\{P\}]\Big[\!\int_{-\pi}^\pi\!\!\!\rmd\theta~P(\theta)f(\theta)\!\Big]\Big[\!\int_{-\pi}^\pi\!\!\!\rmd\theta~P(\theta)g(\theta)\!\Big]
\end{eqnarray}
For $T\to\infty$ (i.e. $\beta\to 0$) the only solution of our equations is the trivial P state of fully random phases $\theta_i$:
$W_{F}[\{P\}|\xi]=W[\{P\}|\xi,k]=W[\{P\}]=\prod_\theta\delta[P(\theta)\!-\!(2\pi)^{-1}]$, which solves out equations at any temperature, but will destabilize
at some $T$ in favour of ordered states with (partially) frozen relations between the phases of the oscillators.

\subsection{Continuous phase transitions away from the incoherent state}

To find continuous bifurcations away from the incoherent (P) state one has to carry out a Guzai expansion
\cite{CooSkaPerPerHatWemNik05} of the functional order parameter equations (\ref{eq:RSmodelB}) around the solution $W_F[\{P\}|\xi]=\prod_\theta\delta[P(\theta)\!-\!(2\pi)^{-1}]$. This will involve the modified Bessel functions $I_m(z)$ \cite{Menzel60}. One writes $P(\theta)\!=\!(2\pi)^{-1}\!+\!\Delta(\theta)$
and $W_F[\{P\}|\xi]\!\to\! \tilde{W}[\{\Delta\}|\xi]$, with $\tilde{W}[\{\Delta\}|\xi]=0$ as soon as $\int_{-\pi}^\pi\!\rmd\theta~\Delta(\theta)\neq 0$ and one expands (\ref{eq:RSmodelB}) in $\Delta(\theta)$:
\begin{eqnarray}
\hspace*{-15mm}
\tilde{W}[\{\Delta\}|\xi]&=& \sum_{k\xi^\prime}p(\xi^\prime,k)\frac{k}{\bra k\ket}\int\!\rmd J~ P(J)
\int\!\prod_{\ell<k}\Big[\{\rmd \Delta_\ell\}\tilde{W}[\{\Delta_\ell\}|\xi^\prime]\Big]\nonumber
\\
\hspace*{-15mm}
&&\hspace*{-5mm}
\times
\prod_\theta \delta\left[\Delta(\theta)-\frac{1}{2\pi I_0(\beta J\xi\xi^\prime)}
\sum_{\ell<k}\int\!\rmd\theta^\prime~\rme^{\beta J\xi\xi^\prime\cos(\theta-\theta^\prime)}\Delta_\ell(\theta^\prime)
\right.\nonumber
\\
\hspace*{-15mm}
&&\left.\hspace*{0mm}
-\frac{1}{2}\sum_{\ell\neq \ell^\prime}^{k-1}\int\!\rmd\theta^\prime\Big(\frac{\rme^{\beta J\xi\xi^\prime\cos(\theta-\theta^\prime)}}{I_0(\beta J\xi\xi^\prime)}-1\Big)\Delta_\ell(\theta^\prime)\Delta_{\ell^\prime}(\theta^\prime)+\order(\Delta^3)\right]
\label{eq:guzai}
\end{eqnarray}
We next evaluate functional moments of both sides of this equation. If the first bifurcation away from the P state is of order $\Delta$,
we multiply by $\Delta(\theta)$ and integrate (functionally) over all $\Delta$, leading to an eigenvalue problem for the functions $\Psi_\xi(\theta)=\int\{\rmd \Delta\}\tilde{W}[\{\Delta\}|\xi]\Delta(\theta)$ subject to the constraint $\int_{-\pi}^\pi\rmd\theta~\Psi_\xi(\theta)=0$:
\begin{eqnarray}
\Psi_\xi(\theta)&=& \sum_{\xi^\prime}\gamma(\xi^\prime)\int\!\frac{\rmd J~ P(J)}
{I_0(\beta J\xi\xi^\prime)}
\int_{-\pi}^\pi\!\frac{\rmd\theta^\prime}{2\pi}~\rme^{\beta J\xi\xi^\prime\cos(\theta-\theta^\prime)}\Psi_{\xi^\prime}(\theta^\prime)
\end{eqnarray}
with $\gamma(\xi)$ as defined in (\ref{eq:gammaxi}).
The solutions are of the form $\Psi_\xi(\theta)=\psi(\xi)\rme^{\rmi m\theta}$, with $m\in\{1,2,3,\ldots\}$ and with
 $\psi(\xi)$ to be solved from the eigenvalue equation
\begin{eqnarray}
\hspace*{-15mm}
\order(\Delta)~{\rm bifurcations:}&~~~~~&
\psi(\xi)= \sum_{\xi^\prime} \Big(\int\!\rmd J~ P(J)
\frac{I_m(\beta J\xi\xi^\prime)}{I_0(\beta J\xi\xi^\prime)}\Big)\gamma(\xi^\prime)\psi(\xi^\prime)
\label{eq:Fbif}
\end{eqnarray}
 For $m=1$ the bifurcating state (F) is one where the oscillators synchronize (partly) to a preferred overall phase, whereas for
 $m>1$ the transition is towards a state with non-uniform phase statistics but without global synchronization 
\cite{CooSkaPerPerHatWemNik05}.

If the first bifurcation away from the P state is of order $\Delta^2$ rather than $\Delta$, so $\int\{\rmd \Delta\}\tilde{W}[\{\Delta\}|\xi]\Delta(\theta)=0$,
we multiply (\ref{eq:guzai}) by $\Delta(\theta_1)\Delta(\theta_2)$ and integrate over all functions $\Delta$,
leading to an eigenvalue problem for the function $\Psi_\xi(\theta_1,\theta_2)=\int\{\rmd \Delta\}\tilde{W}[\{\Delta\}|\xi]\Delta(\theta_1)\Delta(\theta_2)$ subject to $\int_{-\pi}^\pi\rmd\theta_1~\Psi_\xi(\theta_1,\theta_2)=\int_{-\pi}^\pi\rmd\theta_2~\Psi_\xi(\theta_1,\theta_2)=0$:
\begin{eqnarray}
\hspace*{-20mm}
\Psi_\xi(\theta_1,\theta_2)&=& \sum_{\xi^\prime}\gamma(\xi^\prime)\!\int\!\!\frac{\rmd J~ P(J)}
{I^2_0(\beta J\xi\xi^\prime)}
\int_{-\pi}^\pi\!\!
\frac{\rmd\theta_1^\prime\rmd\theta_2^\prime}{4\pi^2}~\rme^{\beta J\xi\xi^\prime[\cos(\theta_1-\theta_1^\prime)+\cos(\theta_2-\theta_2^\prime)]}\Psi_{\xi^\prime}(\theta_1^\prime,\theta_2^\prime)
\nonumber
\\[-2mm]
\hspace*{-15mm}&&
\end{eqnarray}
The solutions are of the form  $\Psi_{\xi}(\theta_1,\theta_2)=\psi(\xi) \rme^{\rmi (m_1\theta_1+m_2\theta_2)}$ with $m_{1,2}\in\{1,2,3,\ldots\}$,
representing new states with `frozen' local phase ordering but no global synchronization, i.e. spin-glass type states (SG), each bifurcating when
\begin{eqnarray}
\hspace*{-20mm}
\order(\Delta^2)~{\rm bifurcations:}&~~~&
\psi(\xi) =\sum_{\xi^\prime} \Big(\int\!\rmd J~ P(J)
\frac{I_{m_1}(\beta J \xi\xi^\prime)I_{m_2}(\beta J\xi\xi^\prime)}{I^2_0(\beta J\xi\xi^\prime)}\Big)
\gamma(\xi^\prime)\psi(\xi^\prime)
\nonumber
\\[-2mm]
\hspace*{-20mm}&&
\label{eq:SGbif}
\end{eqnarray}

The right-hand sides of both (\ref{eq:Fbif}) and (\ref{eq:SGbif}) vanish at $\beta=0$, so the transitions correspond to the smallest $\beta$ such that solutions of (\ref{eq:Fbif}) and (\ref{eq:SGbif}) exist.
Hence we need the maxima of the right-hand sides over $m$ and $(m_1,m_2)$, respectively. The properties of the modified Bessel functions
(see e.g. \cite{CooSkaPerPerHatWemNik05}) ensure that these maxima are found for $m=1$ and $(m_1,m_2)=(1,1)$.
Finally, if we again choose  the bond distribution
$P(J)=\frac{1}{2}(1\!+\!\eta)\delta(J\!-\!J_0)+\frac{1}{2}(1\!-\!\eta)\delta(J\!+\!J_0)$,  the bifurcation conditions can once more be written in the form (\ref{eq:PtoF},\ref{eq:PtoSG}),
but where in the case of coupled oscillators the largest eigenvalues $\lambda_{\rm max}^{(1)}$ and $\lambda_{\rm max}^{(2)}$
refer to the following matrices
\begin{eqnarray}
&&
M^{(1)}_{\xi\xi^\prime}(\beta)= \eta\frac{I_1(\beta J_0\xi\xi^\prime)}{I_0(\beta J\xi\xi^\prime)}\gamma(\xi^\prime),
~~~~~
M^{(2)}_{\xi\xi^\prime}(\beta)= \frac{I^2_{1}(\beta J_0 \xi\xi^\prime)}{I^2_0(\beta J_0\xi\xi^\prime)}\gamma(\xi^\prime)
\label{eq:MmatricesB}
\end{eqnarray}
Comparison with (\ref{eq:MmatricesA}) shows that, inasmuch as the location of the transition lines away from the P state is concerned,
the differences between having interacting Ising spins or coupled oscillators on the nodes of the network
are accounted for by the simple substitution $\tanh(z)\to I_1(z)/I_0(z)$ in the relevant remaining eigenvalue problem.

\end{document}